# Observation of poloidal magnetic flux emission from a low-pressure spark: validation of the hypothesis of constrained plasma dynamics?

**A B Blagoev[1,2], V Yordanov[1] and S K H Auluck[2]**

1. University of Sofia, Faculty of Physics, 5 J. Bourchier blvd., 1164, Sofia, Bulgaria
2. International Scientific Committee on Dense Magnetized Plasmas, Hery 23, P.O. Box 49, 00-908 Warsaw, Poland

1. Corresponding author: blagoev@phys.uni-sofia.bg

Abstract:

This paper revisits an earlier discussion of the hypothesis of constrained dynamics [IEEE Trans. Plasma Sci. **41**, 2013 pp. 437-446] with better experiments and insights. This hypothesis pertains to the observation that the standard single fluid equation of motion of a current-carrying plasma has terms governed by diverse physical phenomena which are constrained to evolve at widely mismatched characteristic space and time scales. It should then be possible to construct counterexamples where the equation cannot be satisfied at all times. The hypothesis posits that the usually neglected electron inertia terms come into play in such examples and create "anomalous" electron currents whose signatures can be detected outside the plasma. One of these signatures is emission of poloidal magnetic flux not attributable to a helical deformation of the current channel. Following the cue of the above-mentioned paper, a series of low-pressure spark experiments have been performed. Clockwise and counterclockwise diamagnetic loops, an electrostatically shielded magnetic probe oriented to intercept the azimuthal magnetic flux and an electric displacement probe are deployed outside the spark channel. Observation of several expected signatures provides a tentative validation of this hypothesis. This suggests that summary neglect of electron inertia in continuum models of plasma needs to be replaced by a redevelopment of these models based on perturbation theory with an electron-mass-related small parameter. Major insights could then arise into unsolved problems such as the origin of axial magnetic field in astrophysical jets. Near-solid-density magneto-inertial confinement fusion may be a potential practical application.

## I. Introduction

This paper revisits the earlier discussion [1] of the hypothesis of constrained dynamics with better experiments and insights. This hypothesis deals with a feature of the equations that express Newton's Second Law for a multi-component plasma in a single-effective-fluid formalism [2], with inbuilt recognition of the fundamental inseparability of the plasma into its constituent species. The detailed derivation of such equations [2] leaves no doubt that they must be obeyed at all times without exception.

The most commonly used single-fluid model of the plasma, however, omits some terms of such equations which are proportional to electron mass. The resulting equation of motion consists of terms which are constrained to evolve at widely mismatched characteristic space and time scales because they represent diverse independent physical processes. One could then construct experimental counterexamples where the truncated equation of motion cannot be satisfied even in principle. The hypothesis of constrained plasma dynamics posits that in such cases the normally omitted electron inertia terms produce "anomalous" electron currents, creating electric and magnetic fields which instantaneously rectify the imbalance in the standard equation of motion. The postulated electron currents can produce observable signatures in detectors placed outside the plasma.

This paper recapitulates the relevant theoretical background, re-examines this hypothesis in more detail, discusses expected experimental signatures and describes experiments that apparently provide a tentative validation of the hypothesis.

### a. Theoretical background and re-examination of the hypothesis of constrained plasma dynamics

The derivation of the continuum model of plasma [2] treats it as a collection of a large but finite number of electrons, ions and neutral particles. Using methods of statistical physics, it considers the evolution of the probability distribution functions for each of the species comprising the plasma in the 7-dimensional phase space consisting of 3 position coordinates, 3 velocity coordinates and time. If the particles have no interaction among themselves, conservation of the number of each species leads to a homogeneous Boltzmann equation for each species. Taking the velocity moments of these homogeneous Boltzmann equations leads to equations which can be interpreted as conservation laws for mass, momentum and energy of hypothetical fluids – an electron fluid, an ion fluid and a neutral fluid. Each of these hypothetical fluids is said to move with its own fluid velocity, which is defined in terms of the first velocity moment of its distribution function. Each fluid has a spatially distributed number density defined by the zeroth velocity moment of its distribution function. The equation resulting from the second velocity moment can be interpreted in terms of an energy conservation law. When interparticle interactions are taken into account, each of the Boltzmann equations becomes inhomogeneous with the appearance of collision operators. The conservation equations for the mass, momentum and energy for different fluids are then coupled via terms which exchange mass, momentum and energy between and within fluids (for example, recombination of electrons and ions reduces numbers of both these species and adds to neutrals and ionization of neutrals does the opposite)

This continuum approach becomes viable only if the hierarchy of moment equations can be truncated. For this, it is necessary that the higher order moments be finite and expressible in terms of the lower order moments [2] – a major condition that translates into restrictions on the applicable plasma parameter space. The considerable practical utility of the continuum approach is also obtained at the cost of ignoring important physical phenomena such as wave-particle interactions.

Inseparability of the three hypothetical fluids is a fundamental physical property of the plasma. This property is built into the continuum description of the plasma by reducing the many-fluid description to a single-effective-fluid description [2]. For this purpose, an average single-fluid velocity is defined as the velocity of the centre of mass of the three (or even more) fluids [2,3]. The difference between the velocities of the ion and electron fluids gives rise to an effective electric current density, which becomes a source term in Ampere's Law [2,3]. The effective mass density is computed as the weighted average of the mass densities of the multiple fluids. The difference in the spatial distribution of number densities of the ion and electron fluids gives rise to a spatially distributed charge density, which becomes a source term in Gauss' Law [2]. The many-fluid equations can be transformed into single-fluid equations by performing suitable mathematical operations [2,3] which express the many-fluid variables in terms of the single-fluid velocity, a current density, a stress tensor, a bulk resistivity and so on. These equations (for example, equations 2.200 to 2.210 of Ref [2], with applicable validity conditions) include those popularly known as the equation of continuity, the equation of motion (also known as the momentum equation), the energy equation and the Generalized Ohm's Law, and are supplemented by Maxwell's equations.

The more complete derivation of these equations [2], which fully accounts for the presence of neutrals, covers some aspects (such as the dynamics of neutrals) which are not required for the present discussion. Instead, a more limited treatment [3], that ignores the presence of neutrals but takes into account all the electron-inertia terms, is used below for illustration, with the understanding that in a partially ionized plasma, the motion of ions is coupled with that of the electrons by the long-range Coulomb force and with the motion of neutrals by collisions. The single-fluid equation of motion then becomes [3]

$$\rho\left[\underbrace{\frac{\partial\vec{v}}{\partial t}+\left(\vec{v}\cdot\vec{\nabla}\right)\vec{\nabla}\vec{v}}_{\text{ACCELERATION}}\right]=\underbrace{\vec{J}\times\vec{B}}_{\substack{\text{MAGNETIC}\\\text{FORCE DENSITY}}}-\underbrace{\vec{\nabla}p}_{\substack{\text{PRESSURE}\\\text{GRADIENT}}}-\underbrace{\vec{\nabla}\cdot\left[\frac{\vec{J}\vec{J}}{\varepsilon_0\omega_p^{\ 2}}\right]}_{\substack{\text{ELECTRON MOMENTUM}\\\text{CONVECTION}}} \tag{1}$$

and the Generalized Ohm's Law becomes [3]

$$\vec{E}+\vec{v}\times\vec{B}=\eta J+\frac{1}{en}\left[\vec{J}\times\vec{B}-\vec{\nabla}p_e-\vec{\nabla}\cdot\left(\frac{\vec{J}\vec{J}}{\varepsilon_0\omega_p^{\ 2}}\right)\right]+\frac{1}{\varepsilon_0\omega_p^{\ 2}}\left[\frac{\partial\vec{J}}{\partial t}+\vec{\nabla}\cdot\left(\vec{J}\vec{v}+\vec{v}\vec{J}\right)\right] \tag{2}$$

The usually neglected electron momentum convection term $\left(\vec{v}_e\cdot\vec{\nabla}\right)\vec{v}_e$ takes the form $\vec{\nabla}\left(\vec{J}\vec{J}/\varepsilon_0\omega_p^{\ 2}\right)$ and appears in both the equation of motion for the single fluid and the Generalized Ohm's Law [3]. Terms proportional to $\vec{\nabla}\cdot\left(\vec{J}\vec{v}+\vec{v}\vec{J}\right)$ in (2) represent convection of ion momentum by electrons and electron momentum by ions during their coupled motion.

It can be easily demonstrated [4] that equation (2) along with Maxwell's 3rd and 4th equations describes three modes of oscillation at the plasma frequency. Expressing the current density as

$$\vec{J}=\mu_0^{\ -1}\vec{\nabla}\times\vec{B}-\varepsilon_0\frac{\partial\vec{E}}{\partial t} \tag{3}$$

and noting that the collision frequency is $\nu=\eta ne^2/m_e$, where $\eta$ is the resistivity, equation (2) can be rewritten as

$$\frac{\partial^2\vec{E}}{\partial t^2}+\nu\frac{\partial\vec{E}}{\partial t}+\omega_p{}^2\vec{E}+c^2\vec{\nabla}\times\vec{\nabla}\times\vec{E}=-\frac{e}{\varepsilon_0 m_e}\underbrace{\vec{\nabla}p_e}_{(1)}+\frac{e}{\varepsilon_0 m_e}\left[\underbrace{\vec{J}\times\vec{B}}_{(2)}-\underbrace{\vec{\nabla}\cdot\left(\frac{\vec{J}\vec{J}}{\varepsilon_0\omega_p{}^2}\right)}_{(3)}\right] \tag{4}$$

$$-\omega_p{}^2\underbrace{\vec{v}\times\vec{B}}_{(4)}+\nu c^2\underbrace{\vec{\nabla}\times\vec{B}}_{(5)}+\varepsilon_0{}^{-1}\underbrace{\vec{\nabla}\cdot\left(\vec{J}\vec{v}+\vec{v}\vec{J}\right)}_{(6)}$$

The three components of the electric field $\vec{E}$ all oscillate at the plasma angular frequency $\omega_p$ and represent three independent electron modes. These modes have a damping rate equal to the collision frequency $\nu$ and a scale length equal to $c/\omega_p$. The source terms driving the modes include (1) the electron pressure gradient (2) the Hall electric field (3) the electron momentum convection term (4) the magnetic flux convection term (5) the magnetic flux diffusion term and (6) momentum cross-convection terms. A current-free plasma supports one curl-free mode driven by the electron pressure gradient.

The point to note is that the essential physics contained in Newton's Second Law applied to individual particles, with the Lorentz force acting on charged particles, is completely preserved in these mathematical transformations. They must therefore be obeyed at all times.

The terms of the single fluid equations directly proportional to the electron mass (via $\left(\varepsilon_0\omega_p{}^2\right)^{-1}=m_e/ne^2$ ) are generally omitted from consideration at the outset, using arguments based on their magnitude relative to other prominent terms [2,3,4]. After this neglect, the truncated equation of motion contains terms which can be generically described as (a) an acceleration term (b) a magnetic force density term (c) a thermal force density term and (d) a viscosity term (not shown in equation (1)).

However, these individual terms are governed by highly dissimilar physical phenomena with their own characteristic physical and temporal scales and associated characteristic speeds. The thermal force density (pressure gradient or divergence of the pressure tensor) depends on processes which increase or decrease the number density of each of the species (diffusion, ionization, recombination, charge exchange, compressive flows etc.) and their thermal energy (losses by energy transport and radiative processes, deposition by dissipative processes and work performed in compressive flows); the acceleration term is governed by the compressible hydrodynamics of the fluid (where equation of state models, internal degrees of freedom and thermodynamics play a major role along with generation and propagation of nonlinear shock and rarefaction waves), the viscosity term is dominated by atomic physics of binary and Coulomb collisions. In this situation, one could construct an experiment which forces the magnetic force density $\vec{J} \times \vec{B}$, governed by an external source of current, to oscillate rapidly on too short a time scale for other terms of the truncated momentum equation to comply with Newton's Second Law at all times. The diverse nature of physical phenomena contained in the truncated momentum equation *constrains* the plasma so that it cannot maintain momentum balance in the face of this rapidly varying magnetic force density. This creates a situation where Newton's Second Law expressed in terms of the truncated single fluid momentum equation for a multicomponent plasma might fail. This would implicate the logic leading to the truncated single fluid momentum equation and invalidate the ab initio neglect of electron inertia terms.

The hypothesis of "constrained plasma dynamics" posits that in such cases, the momentum balance is instantaneously restored by the normally neglected electron inertia terms [2,3,4] proportional to $1/\varepsilon_0\omega_p^2$ in equations (1) and (2). To put it differently, such an experiment would repeatedly take the plasma in and out of the region of parameter space where electron inertia terms (proportional to $1/\varepsilon_0\omega_p^2$) are negligible. Summarily neglecting electron inertia is equivalent to assuming that the plasma frequency is infinite at all times. That logic fails when the electron density is so low (for example in the boundary regions) that the plasma period is not negligible on the dominant evolutionary timescale of the plasma.

Electron momentum generation and transport must then occur instantaneously to supply the deficit in momentum balance during such repeated transitions. Associated electron currents must then lead to emission of magnetic flux, which can be detected outside the plasma. These inferences are a logical consequence of the contradiction between two physical requirements implicitly embedded in the derivation of the continuum model of plasma: (1) the fact that Newton's Second Law as applied to the continuum model of plasma must be obeyed at all times (2) all terms in the commonly used truncated single fluid equation of motion are constrained to evolve at widely mismatched characteristic space and time scales of substantially dissimilar physical phenomena.

Heuristically, the origin of these postulated electron currents can be understood as a combined result of two circumstances: (1) the 3 orders of magnitude difference in the masses of ions and electrons and (2) the constraints imposed by charge neutrality [3,4]. An average force, such as that represented by the electron and ion pressure gradients (which must of the same order of magnitude because of temperature equalization due to collisions and quasi-neutrality) acting on both ions and electrons, would tend to accelerate electrons more than ions because of their smaller mass. This would result in a current in the direction of the force. However, charge neutrality must guarantee that all currents are divergence free – they must loop back. That implies that a net force acting on a plasma element must create an electron current perpendicular to the direction of the net force [3,4] giving rise to convective electron acceleration $\left(\vec{v}_e \cdot \vec{\nabla}\right)\vec{v}_e$ in the direction of the applied force in lieu of the kinetic acceleration $\partial_t \vec{v}_e$. When this net force is the radial pinching force corrected for the thermal force, this current must be helical and must therefore emit magnetic flux, as was indeed observed [1].

A microscopic mechanism has been suggested [4] for the generation of these electron currents taking the example of an axially and azimuthally uniform plasma. Any impulsive perturbation of the plasma (for example, one that changes the electron pressure gradient in equation (4)) launches a radial electron mode wherein the electron fluid radially expands and contracts with respect to the ion fluid at the electron plasma frequency $\omega_p$. The restoring force for this radial electron mode arises from Coulomb's Law. This mode gives rise to a radial electron current density $J_{r,\omega p}$ oscillating at the plasma frequency according to equation (3). An arbitrarily weak, zero-frequency axial component of initial magnetic field $B_{z,0}$ (such as the geomagnetic field) couples with the radial current to create an azimuthal component of Hall electric field $E_{\theta,\omega p} \sim -J_{r,\omega p} B_{z,o} / en$ oscillating at the plasma frequency from the second source term in equation (4). This resonantly excites the azimuthal electron mode in equation (4), wherein the electron fluid oscillates azimuthally with respect to the ion fluid at the electron plasma frequency, thereby creating an azimuthal current density $J_{\theta,\omega p}$ oscillating at the plasma frequency. The restoring force for this electron mode arises from Lenz's Law. The current density $J_{\theta,\omega p}$ acts as a source for an oscillating axial magnetic field component $B_{z,\omega p}$. The Hall electric field then acquires a zero-frequency azimuthal component $E_{\theta,0} \sim -\left\langle J_{r,\omega p} B_{z,\omega p} \right\rangle / en$. Collisional damping of the two electron modes ensures that the radial current $J_{r,\omega p}$ and axial magnetic field $B_{z,\omega p}$, both oscillating at the plasma frequency, are never out of phase, leading to a nonzero time-average $\left\langle J_{r,\omega p} B_{z,\omega p} \right\rangle$ which depends on the collision frequency.

The growth rate of the zero-frequency axial magnetic field, proportional to the curl of the zero-frequency azimuthal Hall electric field according to Faraday's Law, thus turns out to be proportional to the zero-frequency axial magnetic field itself. This leads to exponential growth of the original arbitrarily weak but non-zero seed axial magnetic field, highlighting its active participation in a self-reinforcing positive feedback loop.

Note that this is not a simple source-driven magnetic field growth. Curl of the electric field given by the Generalized Ohm's Law (2) contains the well-known source term $-\vec{\nabla}\times\left(\vec{\nabla}p_e/en_e\right)$ which causes a linear growth of the magnetic field. This growth does not need an initial seed magnetic field and the maximum magnetic field is limited by diffusion of the magnetic flux. Also, this source term is proportional to the vector product of the gradients of density and pressure. For generating the axial component of the magnetic field, at least one of the gradients would have to be in the azimuthal direction, which would imply a deviation from azimuthal symmetry. The growth would in fact be proportional to the deviation from azimuthal symmetry. The mechanism described above operates even in strictly symmetric conditions.

An analytical discussion of this growth is given elsewhere [4]. This growth is halted [5] when a quasistatic equilibrium is established between the magnetic force density $J_\theta B_z$ created by the axial magnetic field and the net rate of "momentum balance deficit". This quasistatic equilibrium condition eliminates the impulsive perturbation which is the source of the radial charge density oscillations in the above mechanism. Essentially, the right-hand side of the radial component of equation (4) becomes zero, which leaves the electron modes to ring down to extinction, halting further growth. This condition can be used to determine the scale and evolution of the generated axial magnetic field on a timescale comparable with the hydrodynamic evolution time scale, which is much larger than the collisional timescale. It would then appear that the plasma evolution is accompanied by an axial magnetic field of unknown origin. Theories of astrophysical jets [6,7] do indeed postulate the existence of an axial magnetic field in an ideally symmetric condition whose origin remains unclear.

The origin of this equilibrium can also be understood in terms a two-dimensional particle orbit argument [5]. A charged particle placed within an arbitrarily weak axial magnetic field and acted upon by a centrally directed force in addition to the Lorentz force moves in such a manner that its orbital current is paramagnetic: it strengthens the initial axial magnetic field. This proposition can be proved algebraically [5] and can also be easily verified by direct numerical computations. The argument fails in the absence of a centrally directed force because the charged particle simply flies away without creating a paramagnetic azimuthal drift current. The expression for the azimuthal drift current leads to the result that the centrally directed force must balance the magnetic pressure exerted by the axial magnetic field. The correspondence between this argument and the nonlinear electron mode interaction mechanism is established when one recognizes that the rate of radial momentum balance deficit per charged particle in a continuum model is identical with the average centrally directed force on a charged particle in the orbit theory argument.

The above discussion suggests that the postulated phenomenon should occur whenever a finite plasma of limited lifetime is created from a no-plasma state by a process that applies a strong impulsive force. There is indeed a recent observation [8] wherein an ultra-intense femtosecond laser pulse interacting with a neodymium magnet having a kilogauss level magnetic field at its surface creates a 10's of mega-gauss magnetic field lasting several picoseconds reporting a $10^4$-fold amplification. Particle-in-cell simulations for a simplified problem formulation corroborate [8] occurrence of an exponential growth of the axial magnetic field.

**b. Experimental motivation and expected signatures:**

The above discussion raises a question mark about the internal self-consistency of the widely used single fluid model of the plasma, particularly in the context of spatially and temporally finite plasmas. However, such discussions are inherently incapable of evaluating the practical significance of such observations. That lies in the domain of experimental research and provides the motivation for considering whether this issue can be demonstrated reproducibly. Such demonstration would then make a case for redevelopment of the continuum model of plasma that does not summarily reject electron inertia. This is discussed in greater detail further in this section.

The above theoretical discussion also suggests the following basis for designing such experiments.

1. Poloidal magnetic flux emission that begins with a rapidly oscillating azimuthal electric field should be an externally observable signature of this phenomenon.
2. Similarly, the radial electron mode causes charge density oscillations which should induce an oscillating charge on any floating conductor nearby and outside the plasma.
3. The onset of electric field oscillations should be correlated with the disappearance of the net centrally directed force.

This paper revisits the experiment described in Ref [1] employing newly demonstrated detectors of magnetic and electric flux emission to observe these signatures. This is essentially a low-pressure spark discharge in air between two conical electrodes connected to a capacitor bank. The current flowing in this discharge has a damped sinusoidal waveform with a time period of a few tens of microseconds. Radial gas dynamic motion of the partially ionized plasma is dominated by the neutrals at room temperature. A current carrying plasma channel resistively dissipating energy is expected to drive a radially diverging shock wave into the surrounding neutral gas. The neutral gas has no rigid boundaries in its vicinity from which to generate a reflected shock wave that can create density oscillations that can follow the oscillating force density. Under this circumstance, an axial magnetic field should be observed in time correlation with the driving magnetic force density. The capacitor discharge current waveform passes through zero multiple times. While approaching the zeroes of the magnetic force density, proportional to the square of the current, the resultant of the pinching magnetic force density and the thermal pressure gradient would stop being a centrally directed force, leading to loss of the expected paramagnetic azimuthal current. As the magnetic force density again increases from its zero level and begins dominating the thermal force density, azimuthal currents are re-established by the mechanism of resonant growth of coupled electron modes described above. This transition should have detectable signatures.

The occurrence of helical currents should lead to emission of poloidal magnetic flux, which can be detected using a pair of clockwise and counterclockwise diamagnetic loops [9]. These helical currents should tend to undergo violent oscillations near the zeroes of the discharge current according to the hypothesis of constrained plasma dynamics described above. The circuit element representation of the discharge plasma [10] should account for the energy consumed and charge diverted in creating these helical currents. Therefore, the emission of electric flux and axial component of current through the discharge should have sharp transients which can be detected using a rate of electric displacement (D-dot) probe [11] and a magnetic probe oriented to intercept the azimuthal magnetic flux.

Observations reported in this paper confirm these expected signatures. These limited data provide at most a tentative validation of the proposed hypothesis, adequate for justifying efforts and expenses for further experimental and theoretical research but perhaps not sufficient to consider the hypothesis as an established fact that can be relied upon uncritically. At the present time, we prefer to retain a question mark in the title over a potential validation of this hypothesis as a reminder that this is still work in progress.

The significance of this exercise is that it invites a redevelopment of fundamental plasma physics in which the electron inertia terms are not summarily rejected on the assumption that they are negligibly small in some region of the plasma. In the boundary regions of a finite plasma and in the initial instants of a plasma being created from a no-plasma state, they cannot be neglected. Any mathematical theory developed using a truncated, zero-electron-inertia formulation of a continuum plasma model cannot use boundary and initial conditions based on a “no plasma state”.  This is because resonant growth of coupled electron modes driven by the plasma formation and sustainment processes would rapidly amplify any ambient magnetic field in the spatiotemporal transition zone between the no-plasma state and the bulk plasma obeying the zero-electron-mass plasma dynamics continuum model. These new fields would invalidate the fundamental assumptions underlying the mathematical theory.

Solutions to many unsolved problems in plasma physics, such as origin of the axial magnetic field in astrophysical jets [6,7], are likely to be discovered in such redeveloped formulation of plasma physics, where a perturbation theory [13] is developed using small parameters involving the electron mass.

This paper is organized into 7 sections. Section II describes the experimental arrangement. Section III reports and comments on the results. Section IV discusses broad features of the results. Section V interprets the results in terms of the hypothesis of constrained plasma dynamics. Section VI elaborates on their significance.  Section VII concludes the paper with a summary.

## II. Experimental setup

This experiment was performed using a modification (see Fig-1) of the Sofia University plasma focus described elsewhere [12]. It has a 20 μF, 40 kV capacitor bank, with a single vacuum switch. It has the usual cylindrical anode surrounded by a squirrel cage cathode. However, the standard 54 mm length glass insulator was replaced with a plastic tube of similar diameter but of 160 mm length. The aim was to prevent formation of a gaseous plasma at the base of the anode. A conical metal extension was fixed to the top of the anode and a similar counter-electrode was positioned above it and joined with the squirrel cage cathode using two extension rods (see Fig-1).

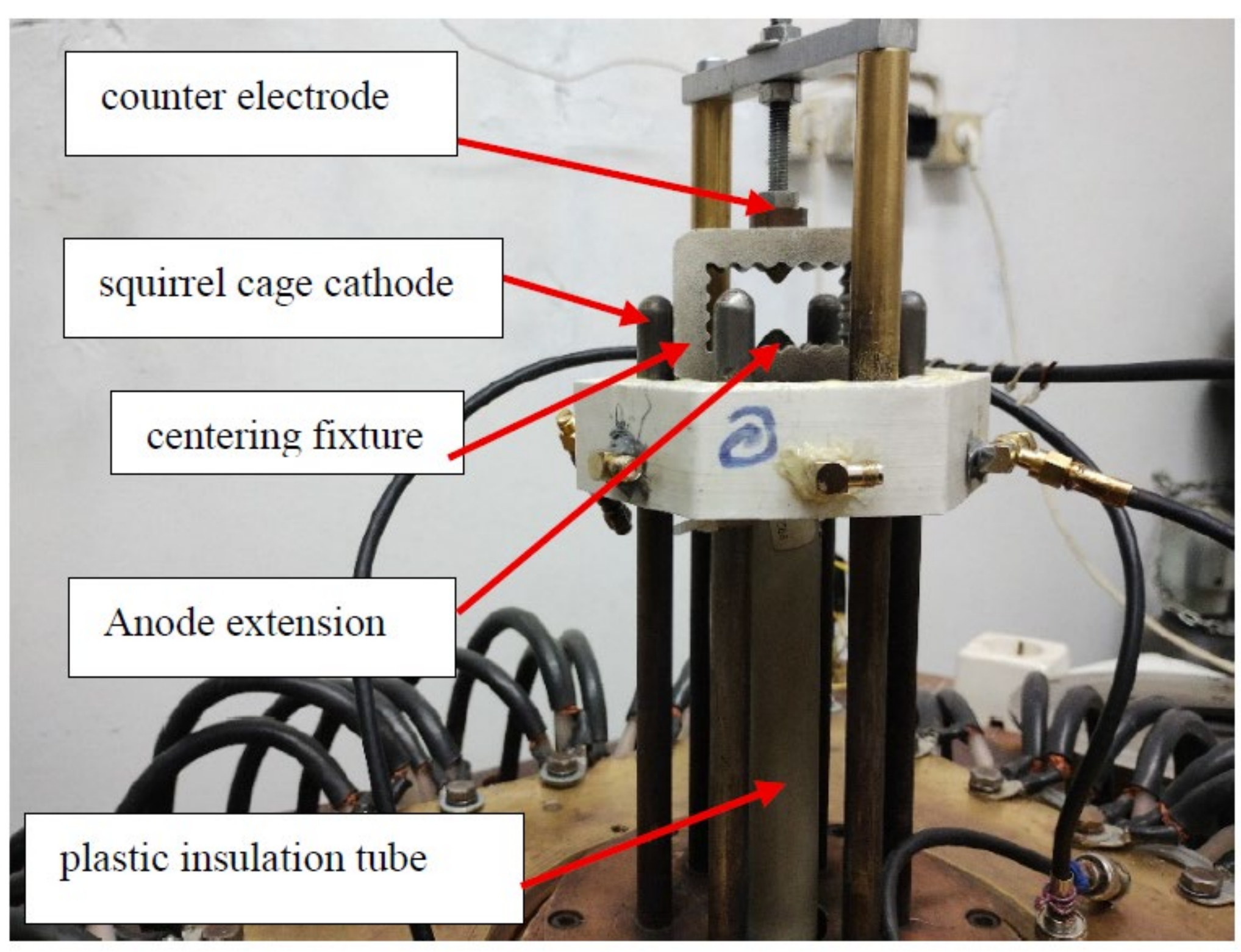

Figure 1(a): Experimental setup

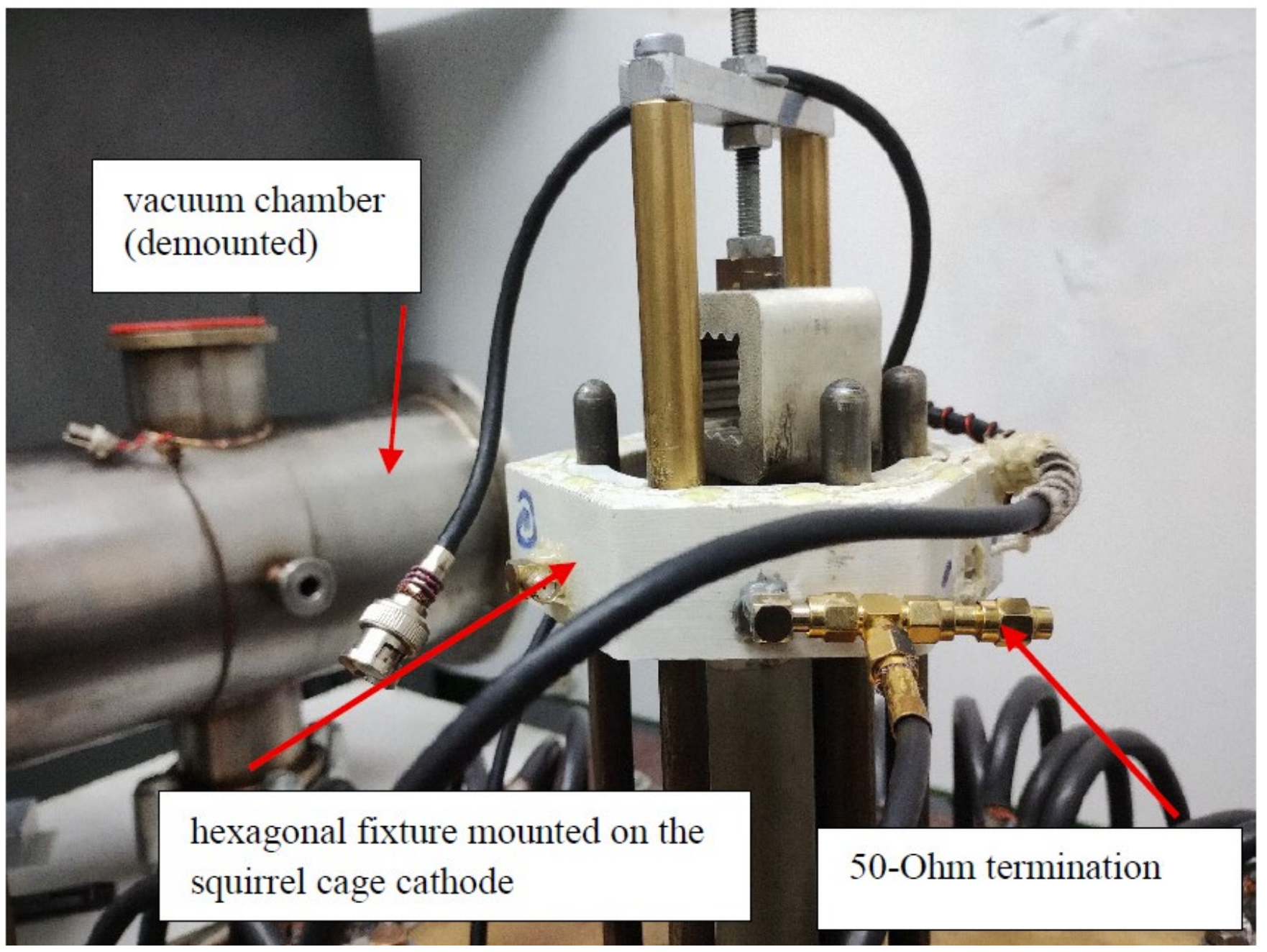

Figure 1(b): Experimental setup: side view

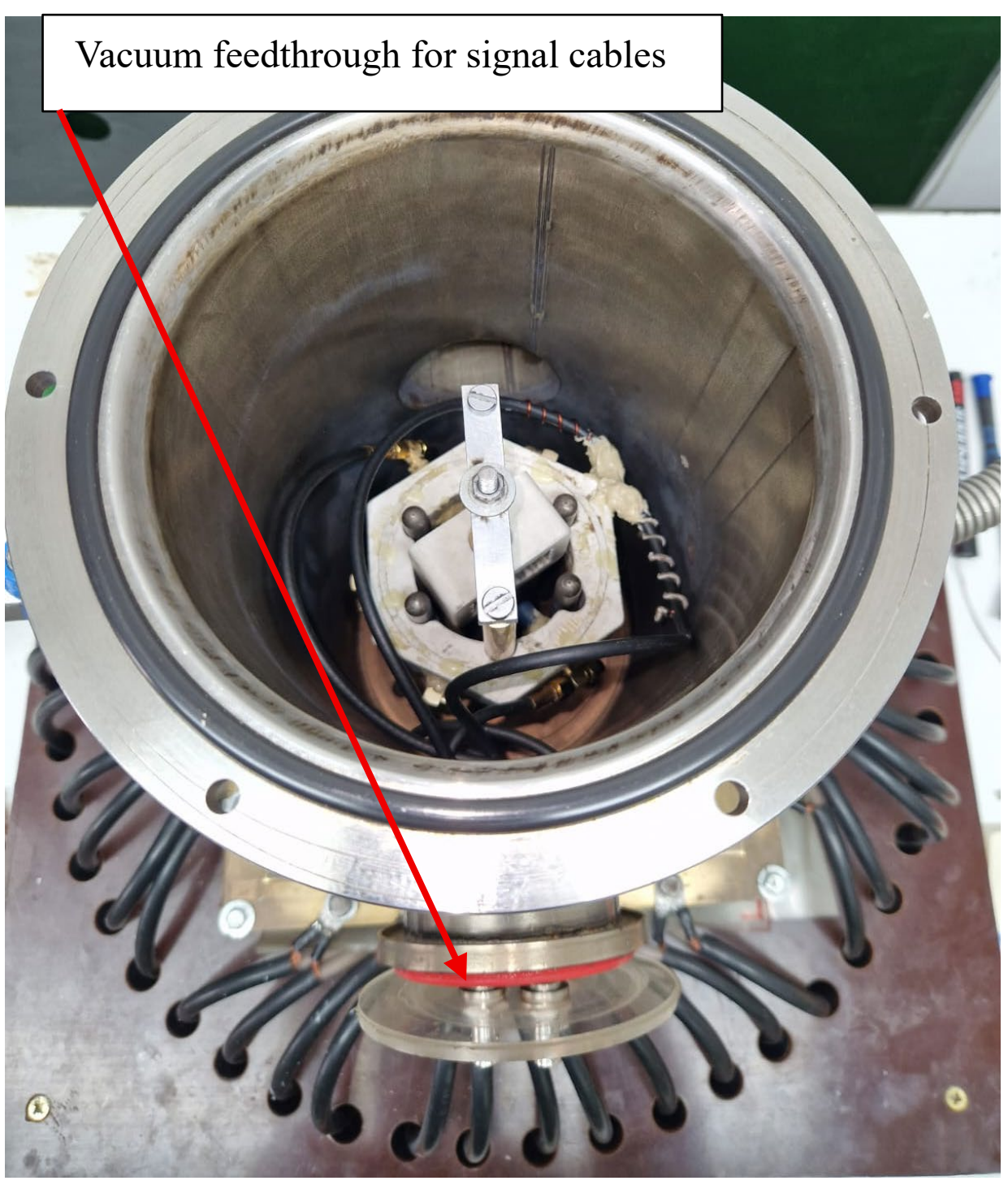

Figure 1(c): Experimental setup: top view

A 3-D printed plastic centering fixture was used to hold the counter-electrode on the same axis as the anode, providing initial symmetry of the discharge plasma. The high voltage switch normally used for plasma focus experiments was bypassed using cables so that the capacitor charging voltage appeared directly across the gap between the pointed ends of the anode extension and the counter-electrode. The charging voltage was gradually increased until electric breakdown occurred across the gap. The vacuum chamber was filled with air at few hundred torr pressure. This arrangement of electrodes increased the inductance of the discharge circuit to nearly 800 nH, enabling many more cycles of current oscillations to observe the zero-crossing events mentioned in the Introduction.

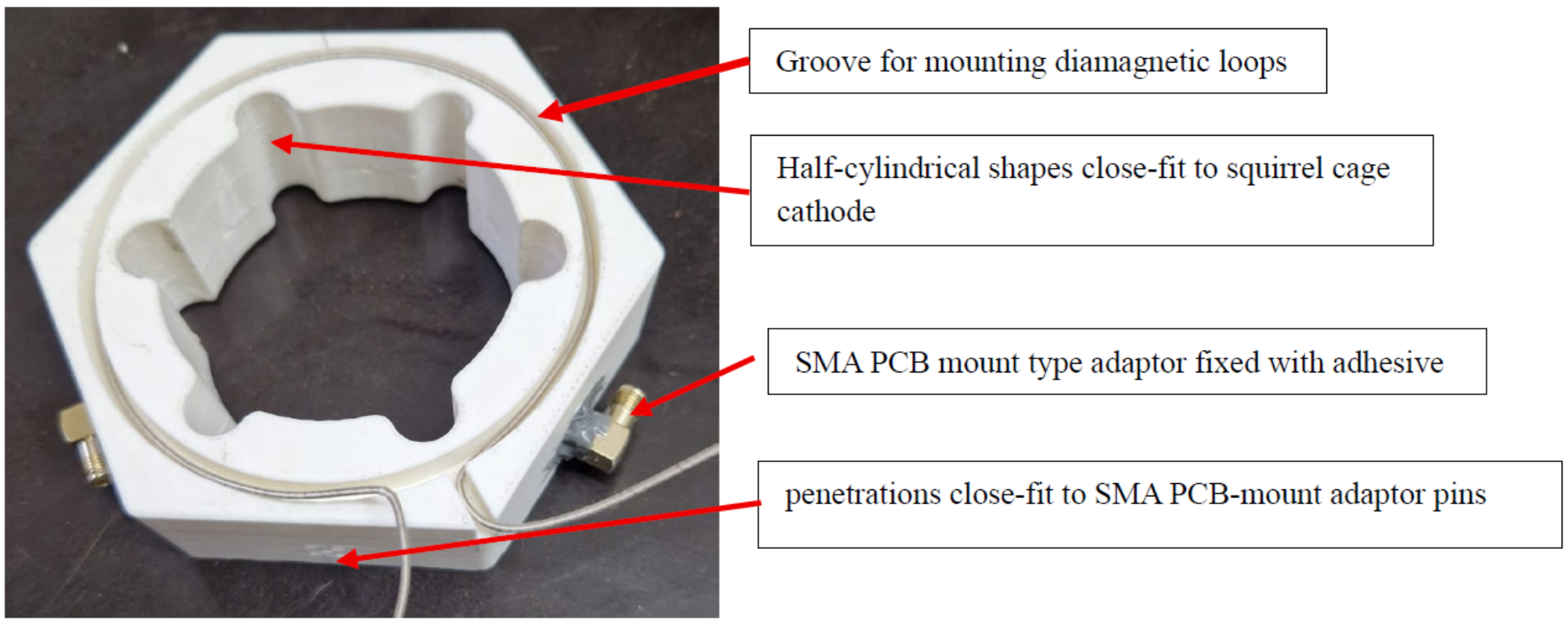

Figure 2: The hexagonal fixture

Another 3-D printed plastic hexagonal fixture (see Fig-2) mounted on the squirrel cage cathode was used to provide support to a pair of diamagnetic loops, a $B_\theta$-dot probe and a D-dot probe [11] (see Fig-3).

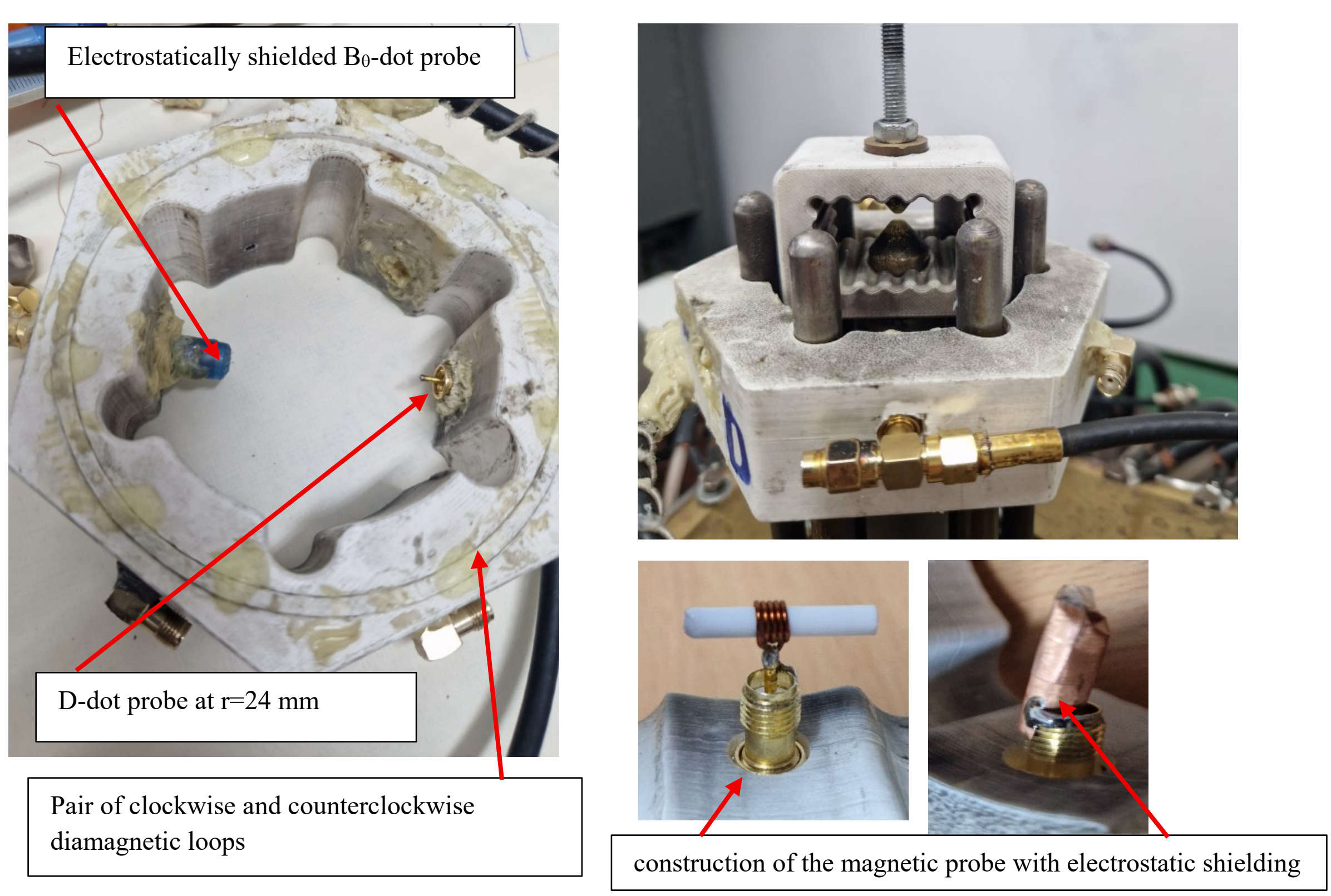

Figure 3. The main diagnostics consists of (1) a $B_\theta$-dot probe (2) a D-dot probe (3) a pair of clockwise and counterclockwise diamagnetic loops supported on the hexagonal fixture mounted on the squirrel cage electrode. The position of the D-dot probe with respect to the mid-point of the interelectrode gap could be adjusted so that it would either face the plasma or the solid electrode. The $B_\theta$-dot probe was made of 5 turns of 0.65 mm diameter wire wound on a 3 mm diameter mandril, subsequently removed. It was electrostatically shielded

The diamagnetic loops were made with 2.5-mm-outer-diameter insulated wire, which were snug fit to the groove in the hexagonal fixture. The centering of the diamagnetic loops with the device axis was ensured by the design of the hexagonal fixture.

The D-dot probe was similar to that described earlier [11] with one essential difference. In the previously reported version [11], its tip was recessed 14 mm inside a 2 mm diameter cavity in the plastic hexagonal fixture. This was intended to reduce the probe's direct contact with the plasma in the rundown phase of the plasma focus. It also provided directional sensitivity to electric-field signals coupled through the cavity, essential for the purpose of that study [11].

In contrast, the probe used in this paper protruded well outside the plastic hexagonal fixture so that it was at a radius of 24 mm.

The final version of the $B_\theta$-dot probe was made using 5 turns of 0.65 mm diameter enamelled copper wire wound on a 3 mm diameter mandril, which was then removed (see Fig-3). Electrostatic shielding was provided using a copper foil covered with transparent adhesive tape ensuring absence of electrically closed paths for eddy currents which could affect the magnetic field measurements. The magnetic probe was placed at a similar radial position as the D-dot probe.

All the signal cables were terminated in the characteristic impedance at both ends. Vacuum compatible BNC female-to-female adapters were used to take the signals outside the vacuum chamber.

The signals were recorded using a 4-channel Tektronix Digital Storage oscilloscope (TDS 3034C).

Experiments were conducted at various positions of the hexagonal fixture, defined in terms of the elevation z of the centre of the D-dot probe from the midplane of the interelectrode gap (defined to be the z=0 plane). At small values of z, the D-dot and the $B_\theta$-dot probes were sensitive to the radial displacement of the plasma while at large values, they faced the solid electrodes and were sensitive only to the circuit-determined variation of the voltage and current. The interelectrode gap and the elevation of the D-dot probes were varied in a spirit of exploration. We present results from a selection of shots which bring out the main points of interest most clearly.

## III. Experimental results:

Selected experimental results are presented in the order they were obtained. Each experimental session led to certain inferences, which influenced the preparations for the next experimental session. As will be shown below (and was also noted earlier [1]), the observed signals have many uncommon and unexpected features. It is natural to ascribe these to deficiencies in the planning, organization and performance of the experiments or to irreproducible artifacts. In order to make a case in favour of the observations being related to real physical phenomena, this section presents data for sets of 3 consecutive shots in several experimental sessions, pointing out noteworthy features and their interrelationships, at the cost of making this an uncomfortably lengthy report.

i. Experiments near 100 torr of air, interelectrode gap ~ 12 mm, D-dot probe facing the plasma

Fig-4 shows data from shot #818, #819 and #820 in air at 96, 98 and 96 torr respectively, with a self-breakdown at 2 kV between electrodes separated by 12 mm. The hexagonal fixture was placed such that the $B_\theta$-dot and D-dot probes were facing the plasma.

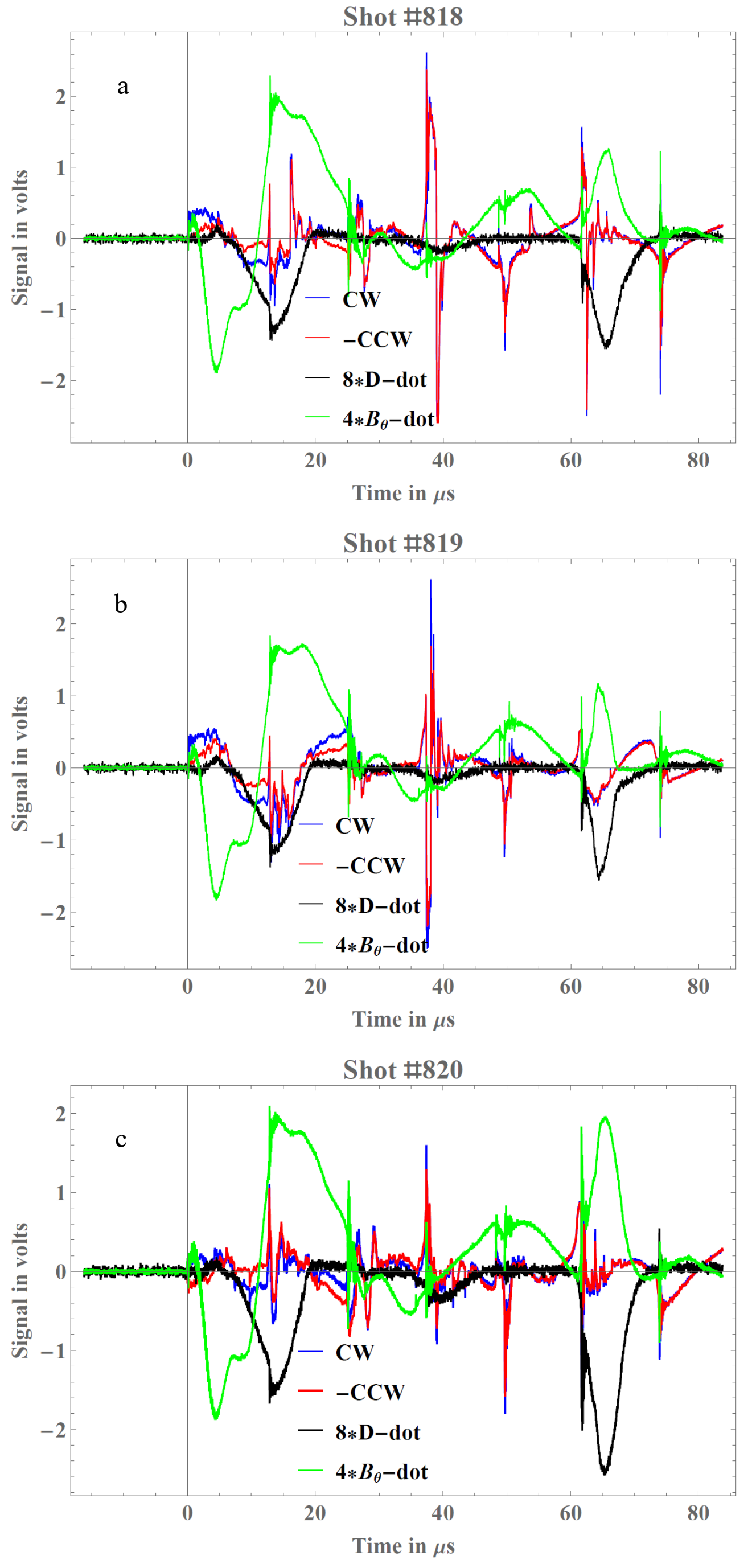

Fig-4. Comparison of data from shot #818, #819 and #820 in air at 96, 98 and 96 torr respectively, with a self-breakdown at 2 kV between electrodes separated by 12 mm. Baseline and time-zero corrections were applied to raw data from each oscilloscope channel. Waveforms recorded at 10 µs/div were smoothed using 3-point moving average method.

Note that the clockwise (CW, along $-\hat{\theta}$) and counterclockwise (CCW, along $+\hat{\theta}$) diamagnetic loop signals are not exact mirror images of each other [9] even though the loops have identical dimensions and nearly identical positions: CW signal differs considerably from the negative of CCW signal as indicated by an imperfect overlap. The CW and CCW signals are superpositions of contributions from the azimuthal electric field induced by the poloidal magnetic flux passing through them and the rate of change of the charge induced on them by the electric flux incident on them [9]. Therefore, half the difference between CW and CCW signals (called the Diff signal) represents the rate of change of poloidal magnetic flux. Similarly, half the sum of CW and CCW signals (called the Sum signal) represents the rate of change of the charge induced on the two loops by the incident electric flux.

The $B_\theta$-dot signal reflects the rate of change of azimuthal magnetic flux passing through the probe. Since the probe faces the plasma, it is sensitive to radial displacements of the plasma.

Similarly, the D-dot signal represents the rate of change of charge induced on a floating conductor by the electric flux incident on it [11]. It is clear from the description in Section II that the electric field geometry is different in the vicinity of the D-dot probe and the diamagnetic loops. The electric flux incident on the plasma-facing D-dot probe is sensitively dependent on the radial displacement of the plasma as well as on the local, circuit-defined voltage on the plasma with respect to the ground potential of the measurement system. The electric charge induced on the diamagnetic loops is not sensitive to radial displacements of the plasma: it is a function only of the circuit-determined voltage.

Fig-5 compares the Diff and Sum signals in reference to the $B_\theta$-dot signal.

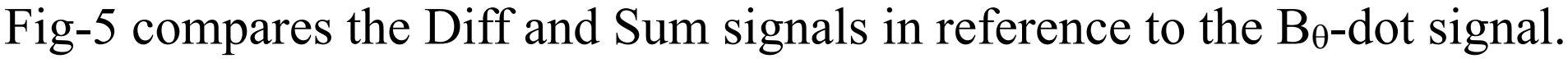

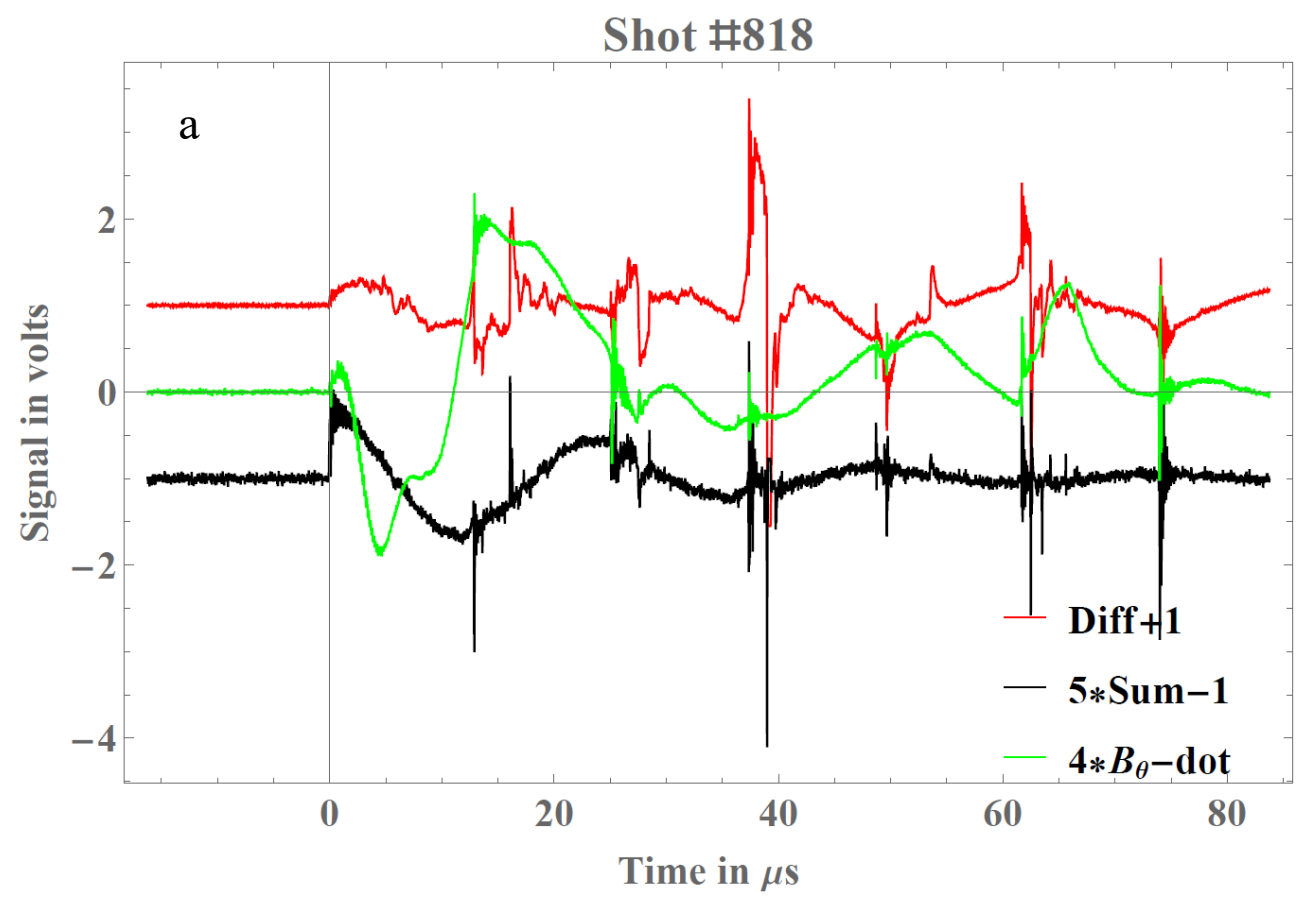

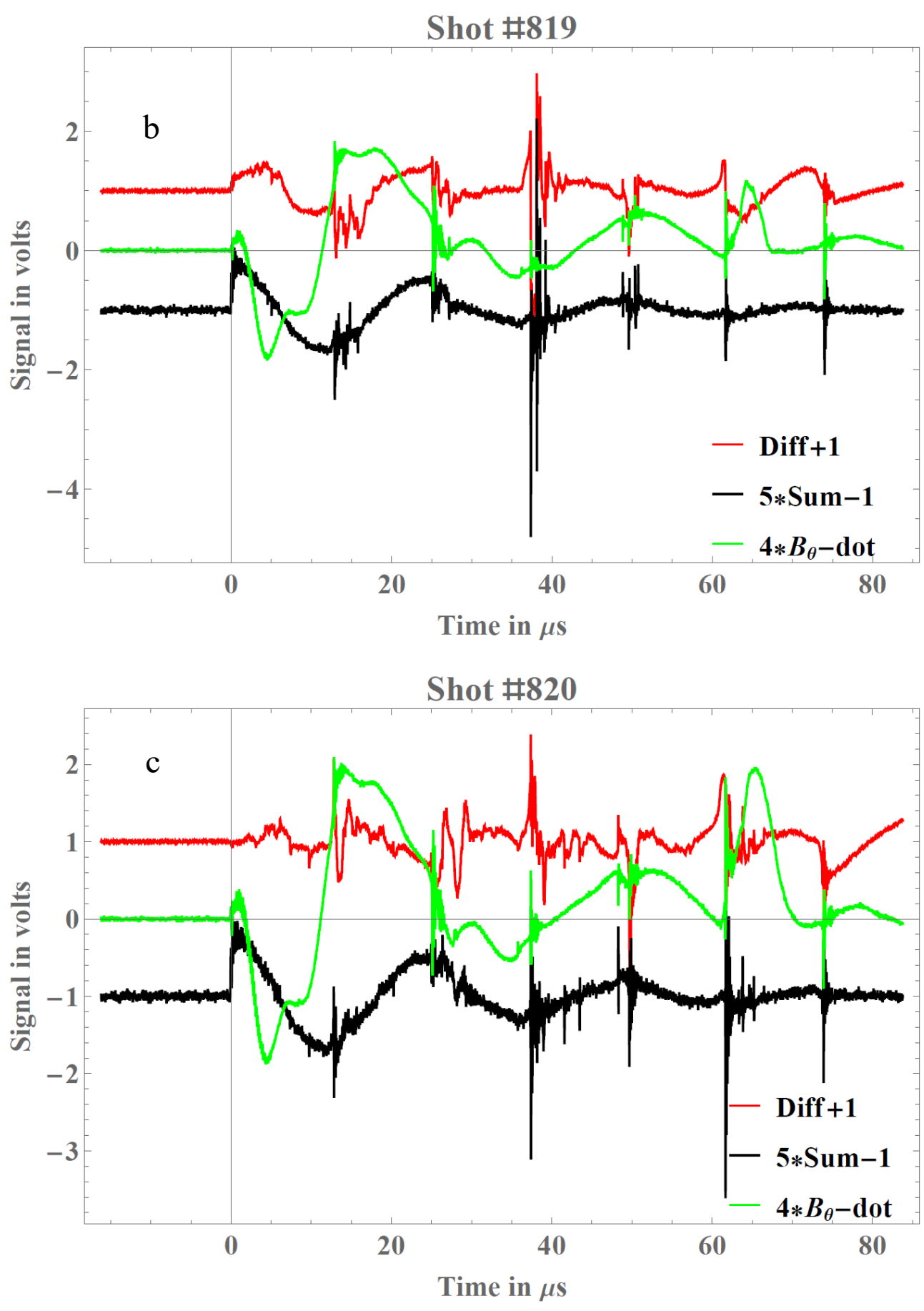

Fig-5: Comparison of Diff=0.5*(CW-CCW) and Sum=0.5*(CW+CCW) signals for the three shots. The signals are offset in order to display features of the two signals without interference. The $B_\theta$-dot signal is shown for reference. Characteristic sharp transients are seen to occur simultaneously in the three signals representing time variation of the emission of poloidal magnetic flux, electric flux and azimuthal (toroidal) magnetic flux.

The strange $B_\theta$-dot waveform, which is proportional to the current derivative and is also a function of the radial position of the plasma, would normally be taken to be a malfunction of the experiment. However, comparison of $B_\theta$-dot signals from the three shots shown in Fig-6 suggests this to be a reproducible feature:

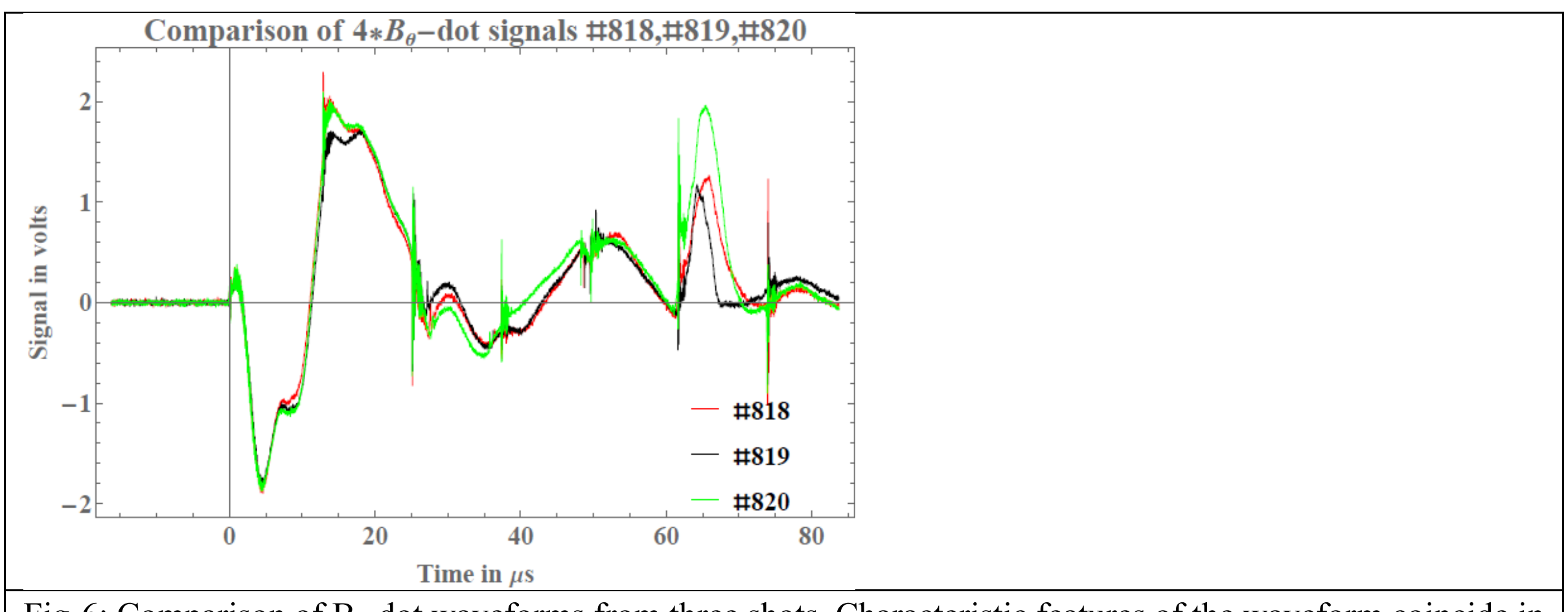

Fig-6: Comparison of $B_\theta$-dot waveforms from three shots. Characteristic features of the waveform coincide in time indicating that they pertain to a reproducible physical phenomenon.

The even stranger D-dot signals [9] are also found to be reproducible as seen in Fig-7.

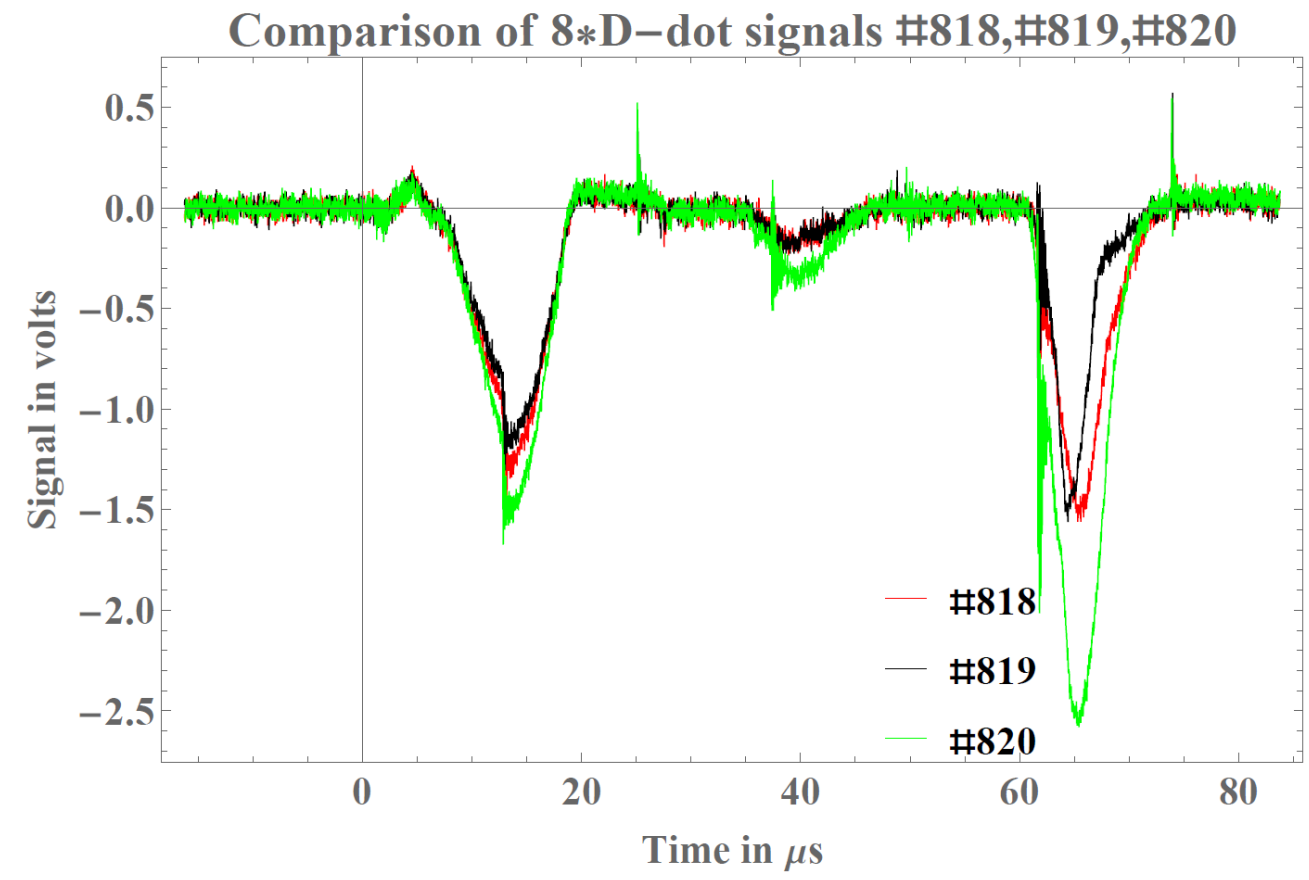

Fig-7: The D-dot signals have two prominent peaks and one smaller peak, all of negative polarity. The shot-to-shot reproducibility is evident.

The D-dot signal may have been influenced by polarization of nearby plastic components in the vicinity of the anode or the probe. In the earlier configuration [11], the probe was designed to face the plasma during the plasma focus formation phase without being immersed in it. These shots followed the same protocol, where the probe tip was recessed 14 mm inside a 2 mm diameter cavity in the plastic hexagon [11]. Its measured response may therefore have been affected by polarization of the surrounding insulator. These possible artifacts were considered in the subsequent series of experiments. Specifically, it led to the D-dot probe being extended outside the hexagon as described in Section II.

It is also likely that the three negative peaks in the D-dot signals represent radial plasma motion: the plasma expands when the magnetic force goes through a zero and again contracts when the magnetic force recovers its strength. It could also be the result of a kink instability.

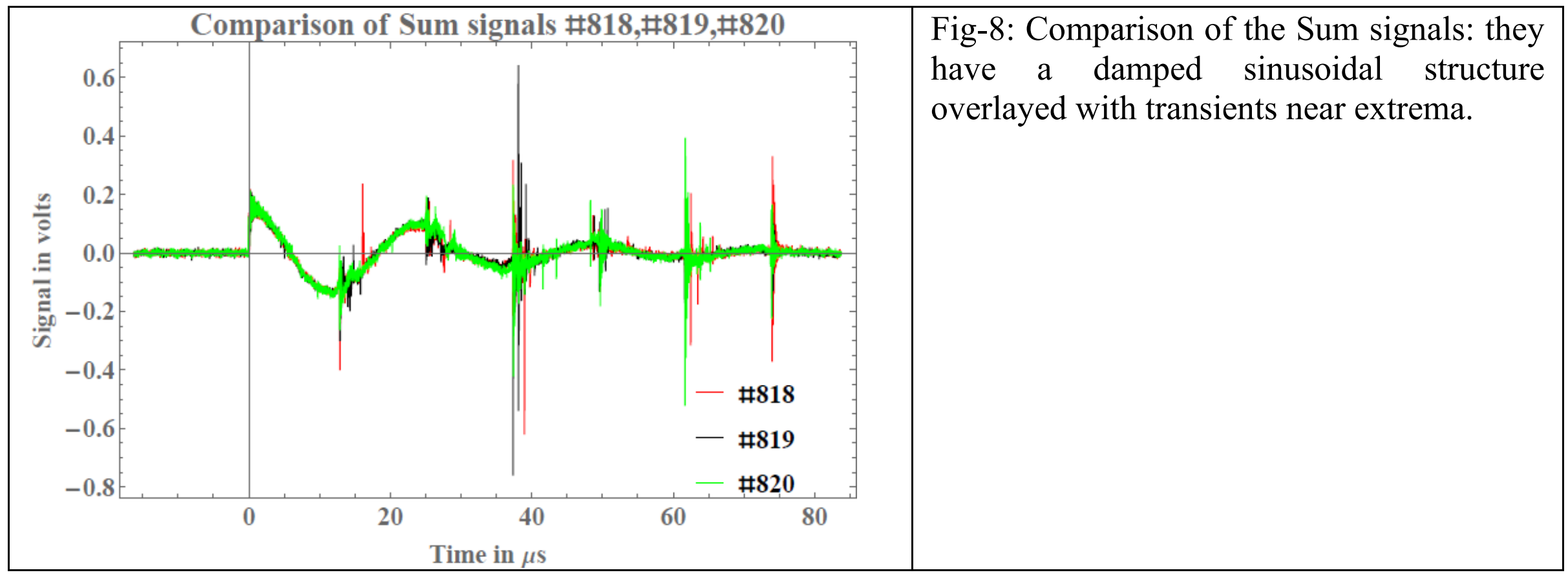

Fig-8: Comparison of the Sum signals: they have a damped sinusoidal structure overlayed with transients near extrema.

Fig-8 shows the comparison of the Sum signals. They show a damped oscillation overlaid with sharp transients.

Fig-9 shows comparison of the Diff signals representing the rate of change of poloidal magnetic flux. It is seen that these signals are certainly not as reproducible as the three discussed above. The transients are time correlated with features in the magnetic probe signal as seen in Fig-5.

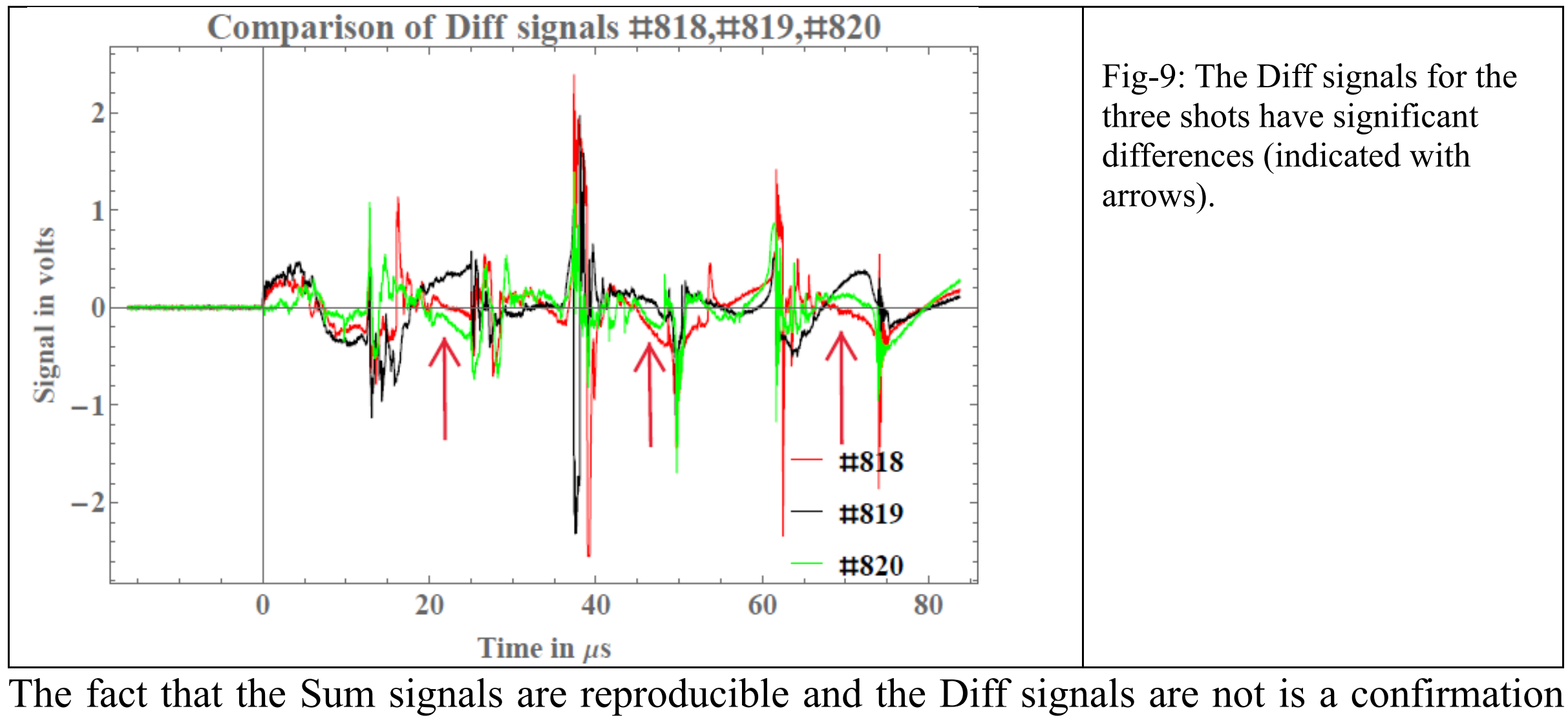

Fig-9: The Diff signals for the three shots have significant differences (indicated with arrows).

The fact that the Sum signals are reproducible and the Diff signals are not is a confirmation that they originate in different physical phenomena: charge induced by the electric flux incident on the diamagnetic loops placed symmetrically about the device axis and electromagnetic induction from the emission of poloidal magnetic flux created by an unknown process.

These differences are reflected in the numerical integration of the Diff and Sum signals shown in Fig-10

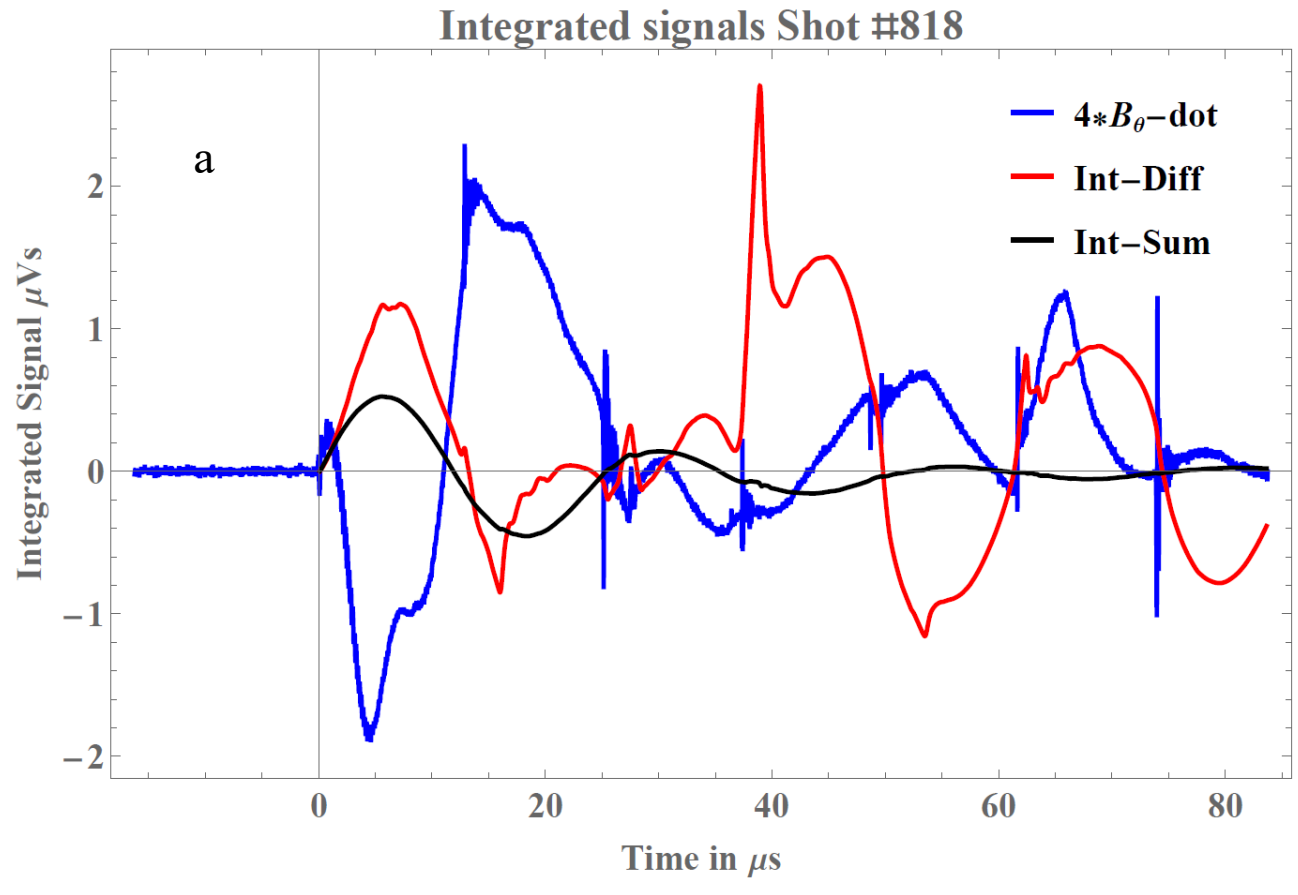

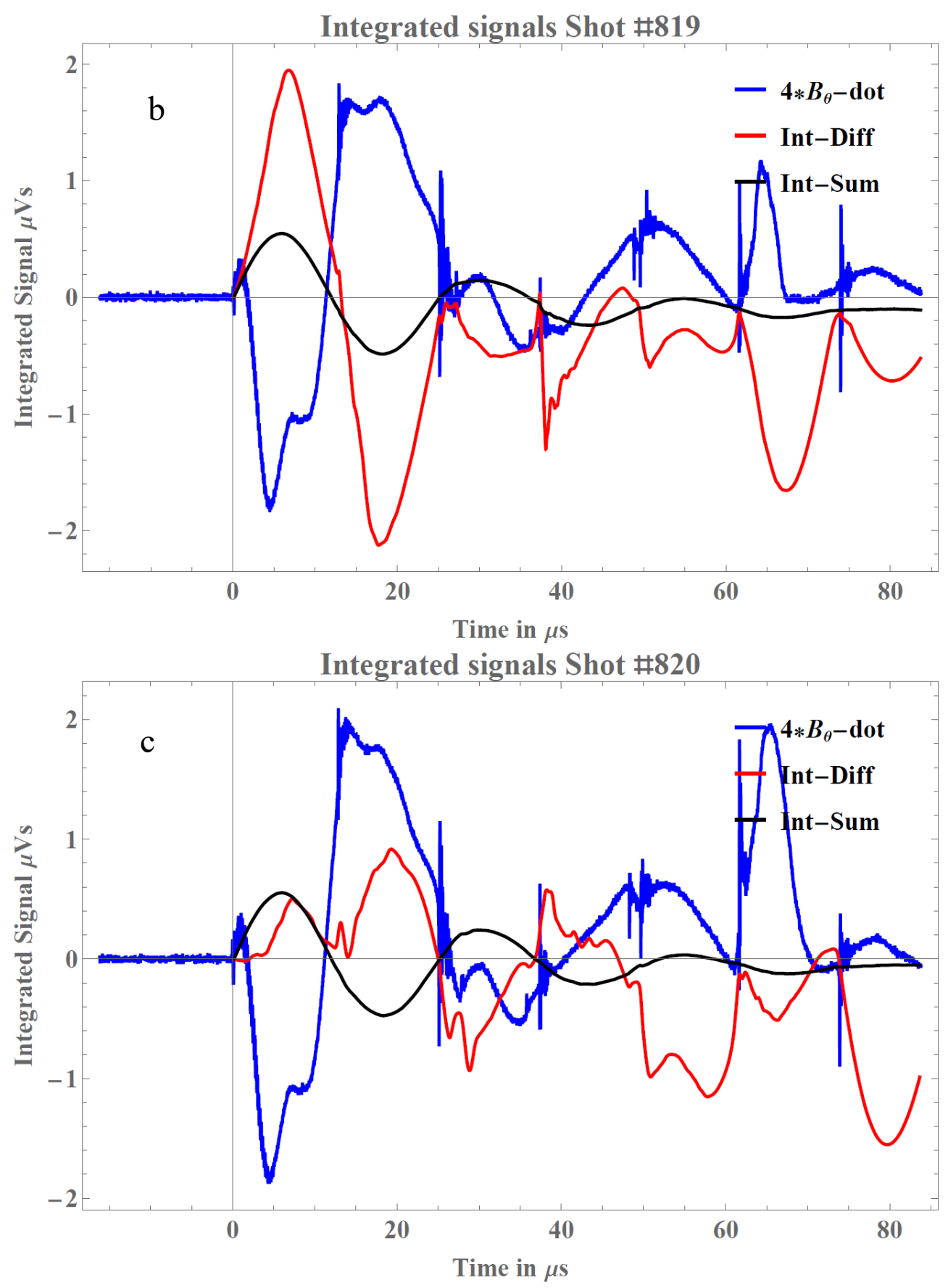

Fig-10: Comparison of integrated Diff and Sum signals for three shots. The $B_\theta$-dot signals are provided as a reference.

Note that the integration of the Sum signal is a damped sine wave, sometimes with a slight inclination of the baseline. This latter feature could either be an artifact of the baseline correction algorithm or could be a real effect attributable to residual charge retention property of energy storage capacitors or some unknown process.

The integration of Diff signal represents the poloidal magnetic flux passing through the diamagnetic loop. The interesting point about these signals in the three shots is the fact that they are radically different in their character. This is in spite of the fact the Sum signals and the $B_\theta$-dot signals are strikingly similar. This implies that some parts of the plasma dynamics are governed by deterministic factors while there are also some chaotic dynamics at play.

ii. Experiments near 200 torr of air, interelectrode gap ~ 3 mm, D-dot probe facing solid electrode

This series of experiments were performed with an interelectrode gap of ~ 3 mm rather than 12 mm as in the previous series because of a suspicion that the strange signals for the $B_\theta$-dot and D-dot probes could be related to a kink instability. Also, the elevation was adjusted so that the D-dot probe faced solid anode rather than the plasma. The magnetic probe was close to the inner surface of the hexagon. The D-dot probe was recessed inside the hexagon. Typical results are shown below:

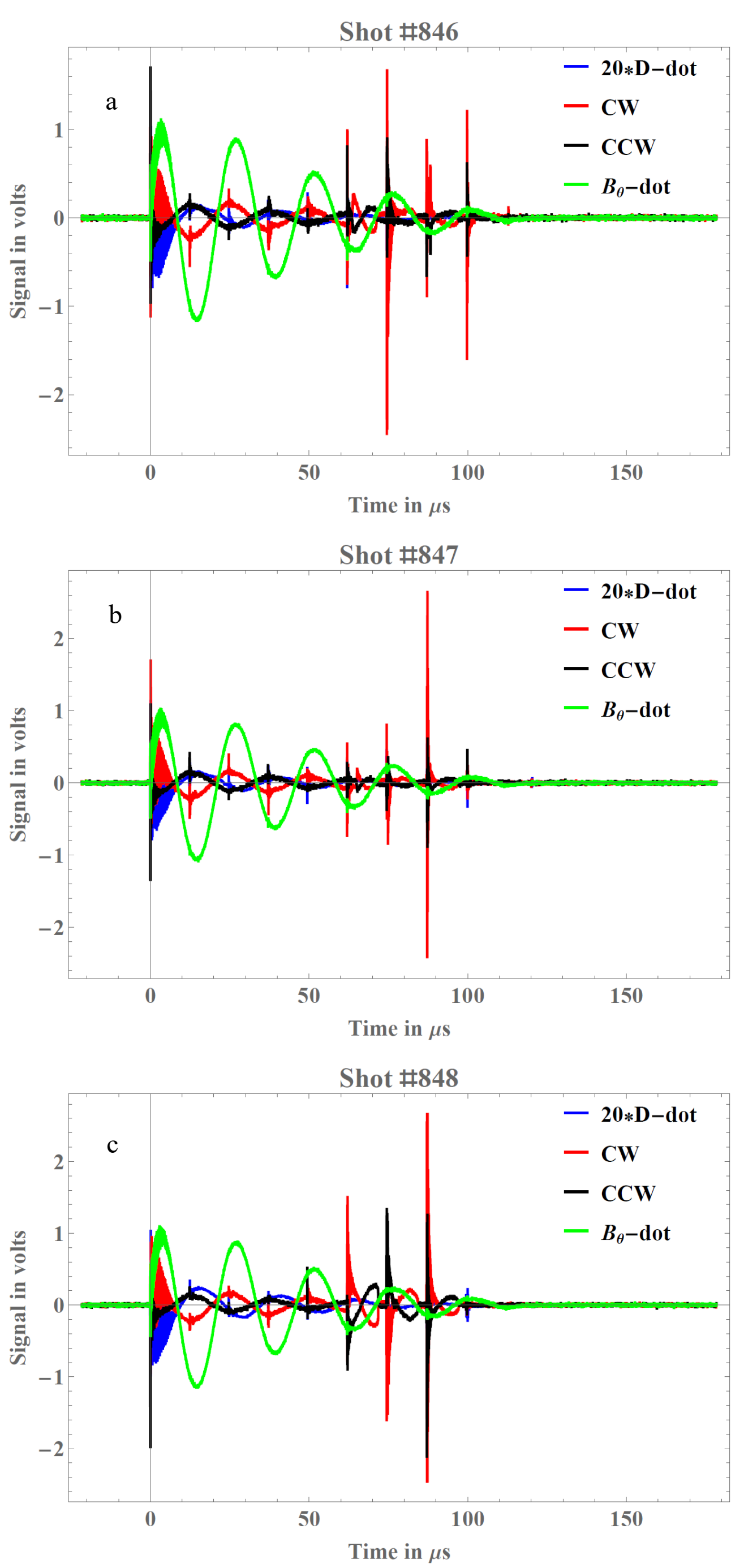

Fig-11: Comparison of three shots at 200 torr of air, breakdown voltage ~2.9 kV. Note the sharp transients after about 3 cycles of current. The oscilloscope was triggered using the $B_\theta$-dot signal. The signals were recorded with a time base of 20 μs/div, with consecutive data points separated by 20 ns. Hence, no smoothing operations were performed. Baseline correction was calculated for each signal as the average of vertical axis values for times 0.1 μs or larger before the trigger instant. Time-zero correction was calculated as the time instant where the $B_\theta$-dot signal exceeded 5 times the magnitude of the baseline correction and applied to all the signals.

Note the fact that all the signals return to their baseline at about 115 μs. The transients in the CW and CCW diamagnetic loop signals are seen to increase towards the later part of the $B_\theta$-dot signal (proportional to the current derivative). There are relatively smaller transients in the $B_\theta$-dot and D-dot signals as well, as seen in Fig-12, the zoomed version of Fig-11.

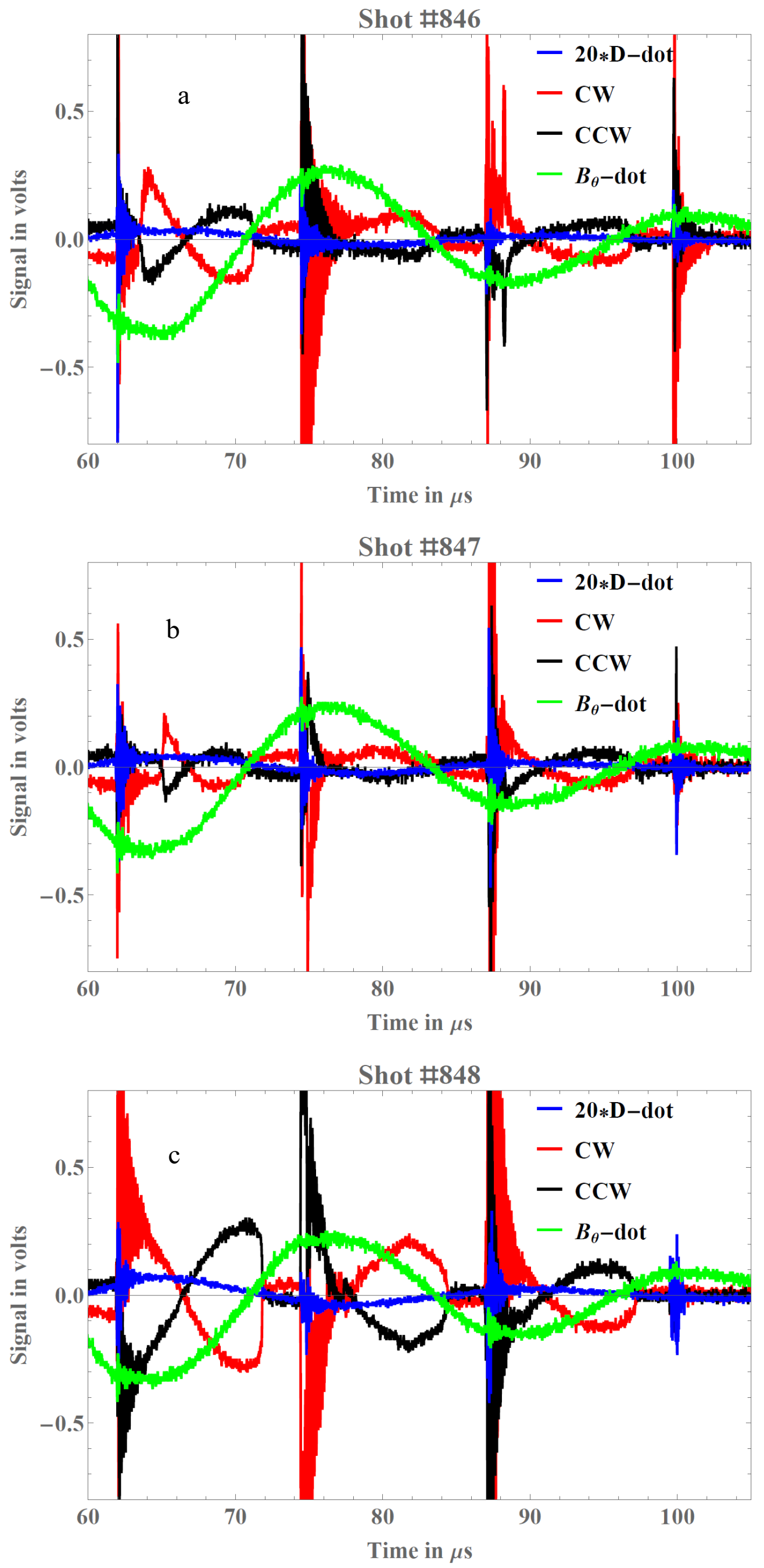

Fig-12: The $B_\theta$-dot and D-dot signals also have transients coinciding with those in the diamagnetic loop signals.

Numerically integrated signals are shown in Fig-13.

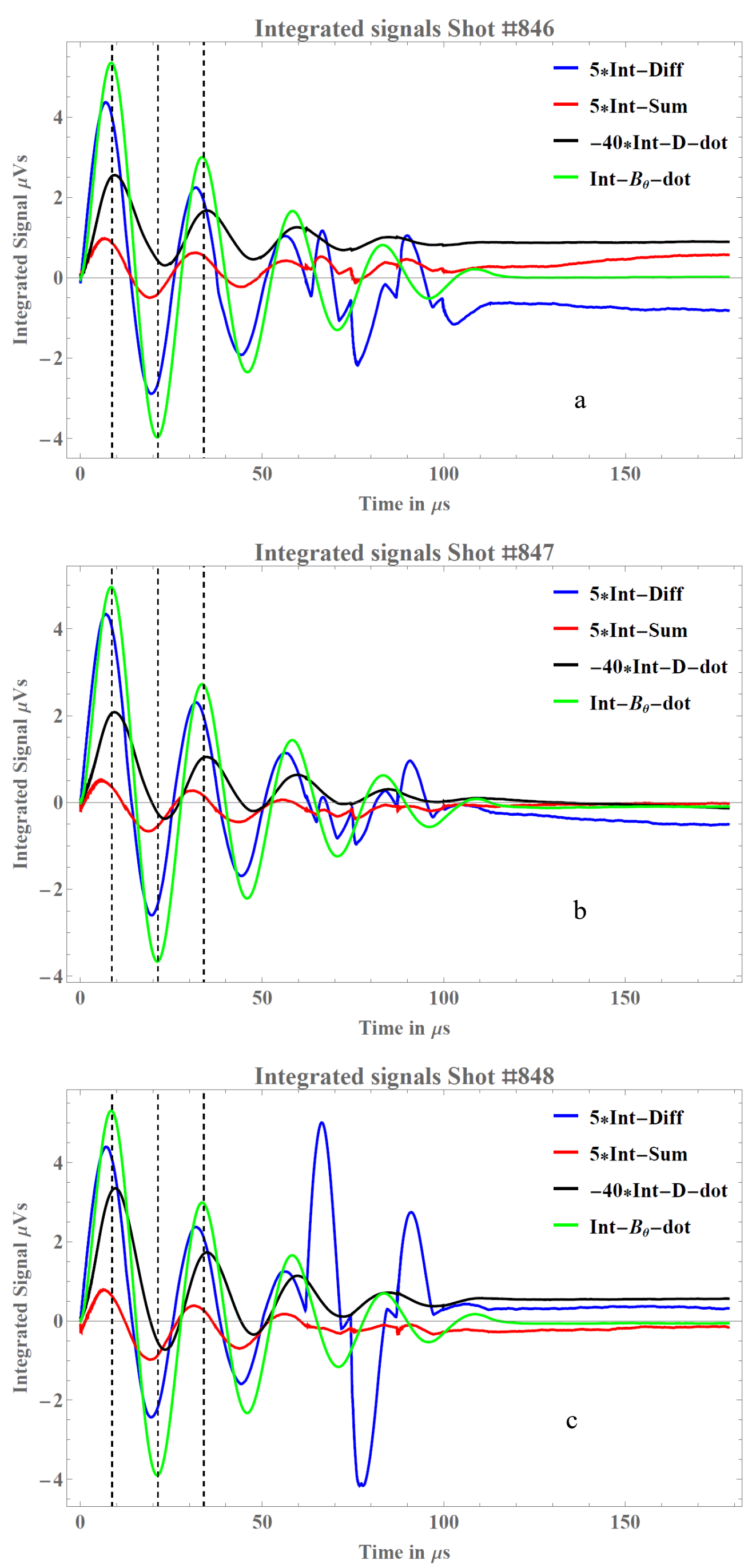

Fig-13: Numerical integration of signals. The vertical dashed lines are drawn through the peaks of the integrated Bθ-dot signal in Shot #846. Note the phase shift of other integrated signals.

This figure exhibits several interesting points.

a. The integration of the Bθ-dot signal is a good damped sinusoidal waveform *that recovers its baseline*. This shows (1) that the algorithm that calculates the baseline, the time-zero corrections and numerical integration works well; (2) the time period is fairly constant ~ 25 μs; (3) the inductance and the resistance of the discharge circuit are in the ranges 792-795 nH and 37-44 mΩ respectively for the three shots; (3) the calibration constant of the integrated $B_\theta$-dot signal is 2.4 kA/μVs, and (4) the amplitude of the damped sinusoidal current waveform is in the range of 14.0-15.5 kA. The significance of this observation lies in several examples shown later where the $B_\theta$-dot signal has transients superimposed on a damped oscillatory waveform and its integration has *a shift in the baseline.*

b. The integrations of the D-dot and Sum signals have a relative phase shift. This phase shift could possibly be due to the plasma being off-axis. The diamagnetic loops are centred on the device axis. The charge induced on them is averaged over the azimuthal angle and is therefore insensitive to radial displacement of the plasma.
c. The integrations of the D-dot and $B_\theta$-dot signals have a small phase shift.
d. Integration of the Diff signal is numerically equal to the poloidal magnetic flux passing through the diamagnetic loops (1 Volt-second=1 Weber).
e. Fig-13 quite clearly demonstrates that the poloidal magnetic flux within the diamagnetic loops is not proportional to the current as expected for a helical deformation of the current channel: the two have strikingly different waveforms. This provides strong support to the notion that there is a mechanism that *creates* poloidal magnetic flux.

The relationship between the transients in the Diff and D-dot signals and the zeros of the current waveform is seen in Fig-14.

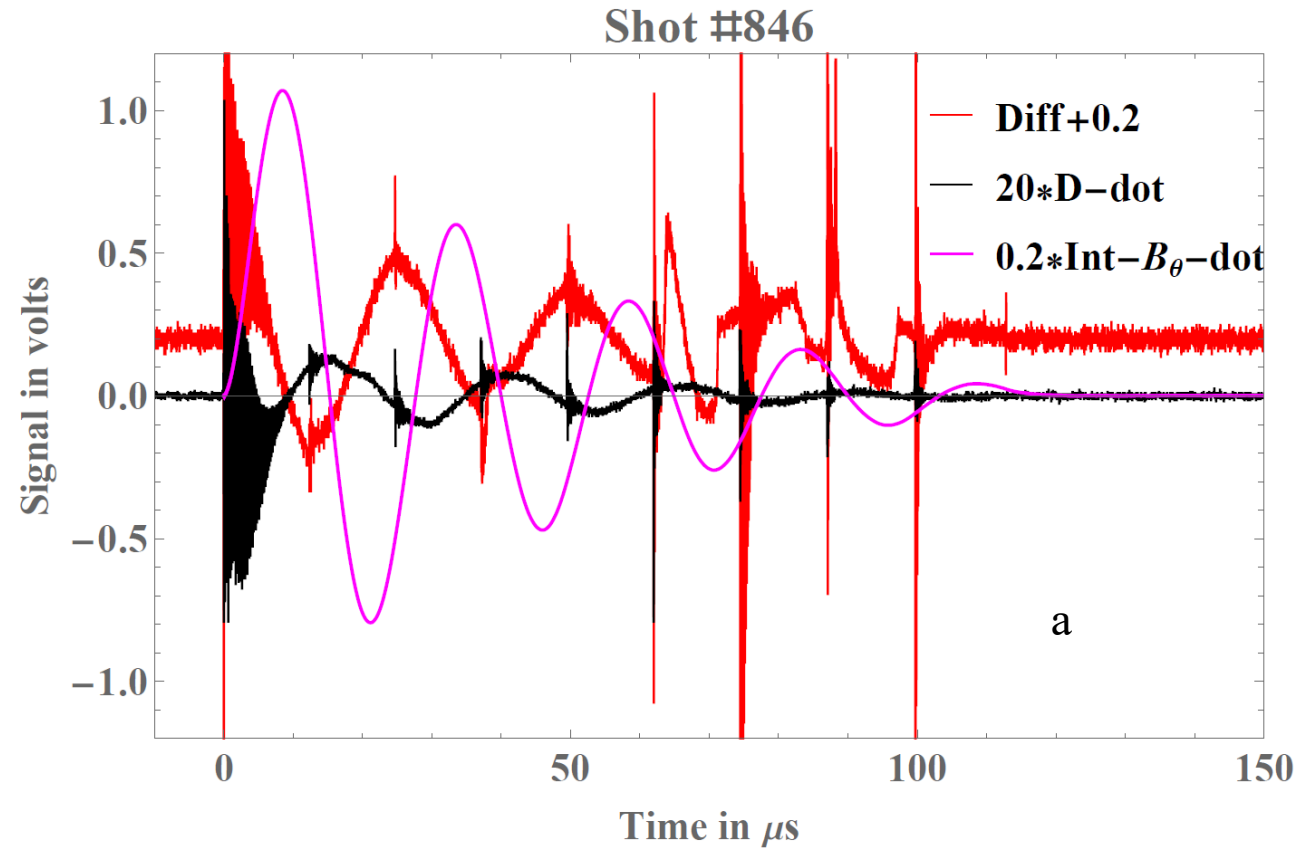

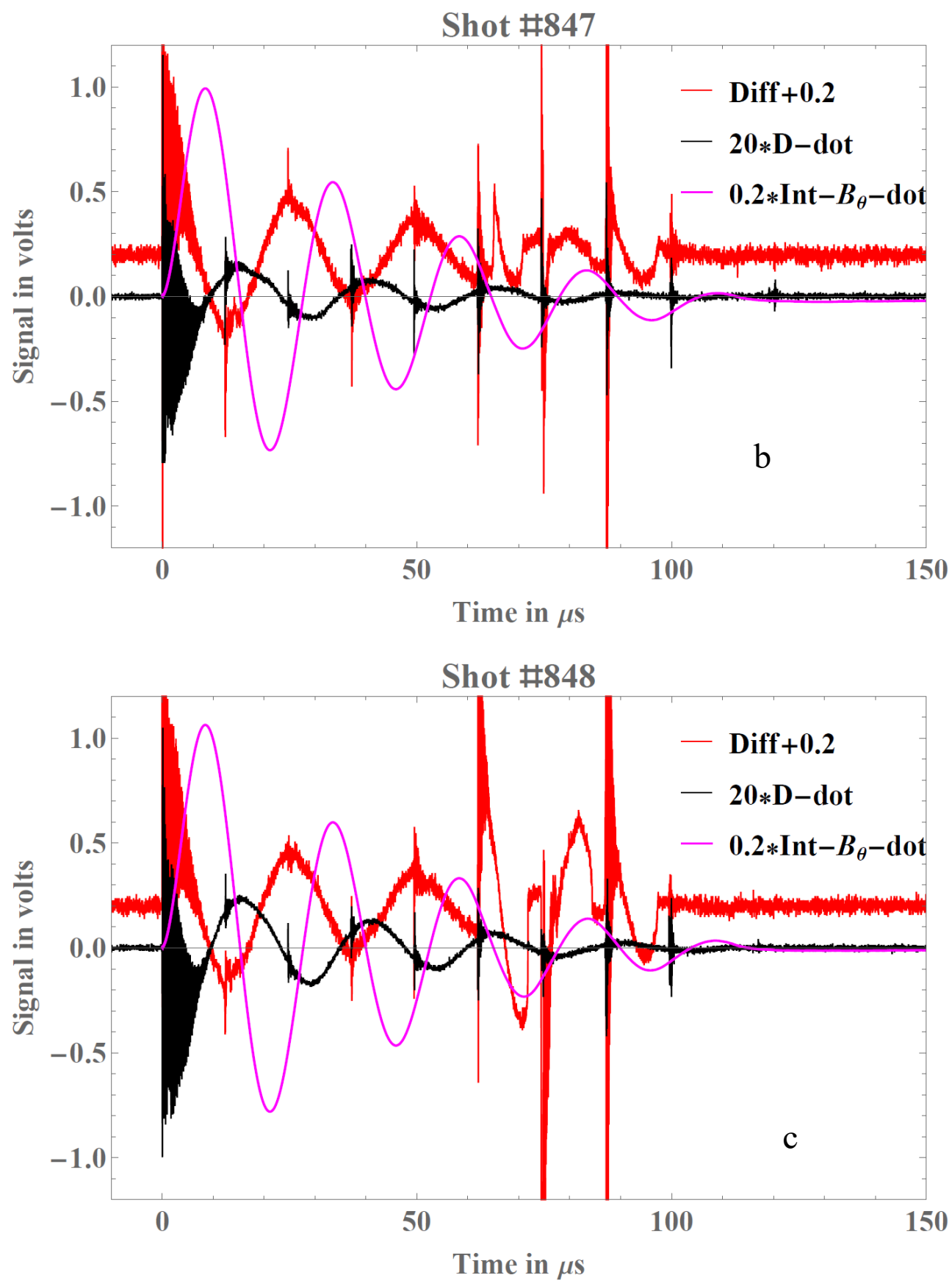

Fig 14: The relationship between the transients in the Diff and D-dot signals and the zeros of the current waveform. The current waveform (integration of $B_\theta$-dot signal) is scaled for display purposes.

Transients in the Diff and D-dot signals are seen to occur consistently before the zeros of the current waveform. One of the striking features of Fig-14 is that the transients in Diff signal increase even as the capacitor discharge current approaches zero.

The detailed nature of the transients in the Diff and D-dot signals and their relation with the zeros of the magnetic force density (square of the integration of the $B_\theta$-dot signal) is seen in Fig-15 and Fig-16.

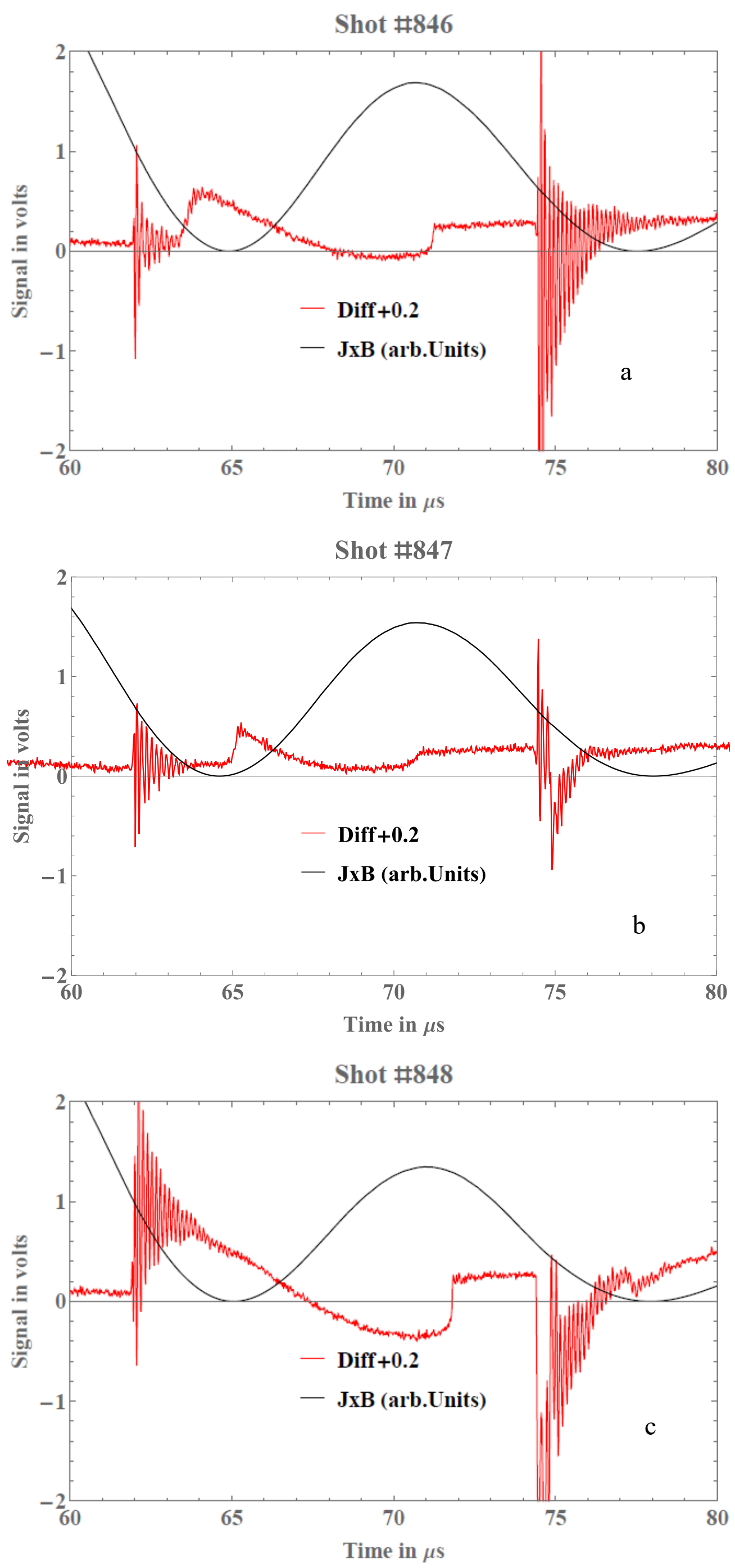

Fig-15: Relationship between the transients in the Diff signal and the zeroes of the magnetic force density. The black curve is the square of the integration of the $B_\theta$-dot signal.

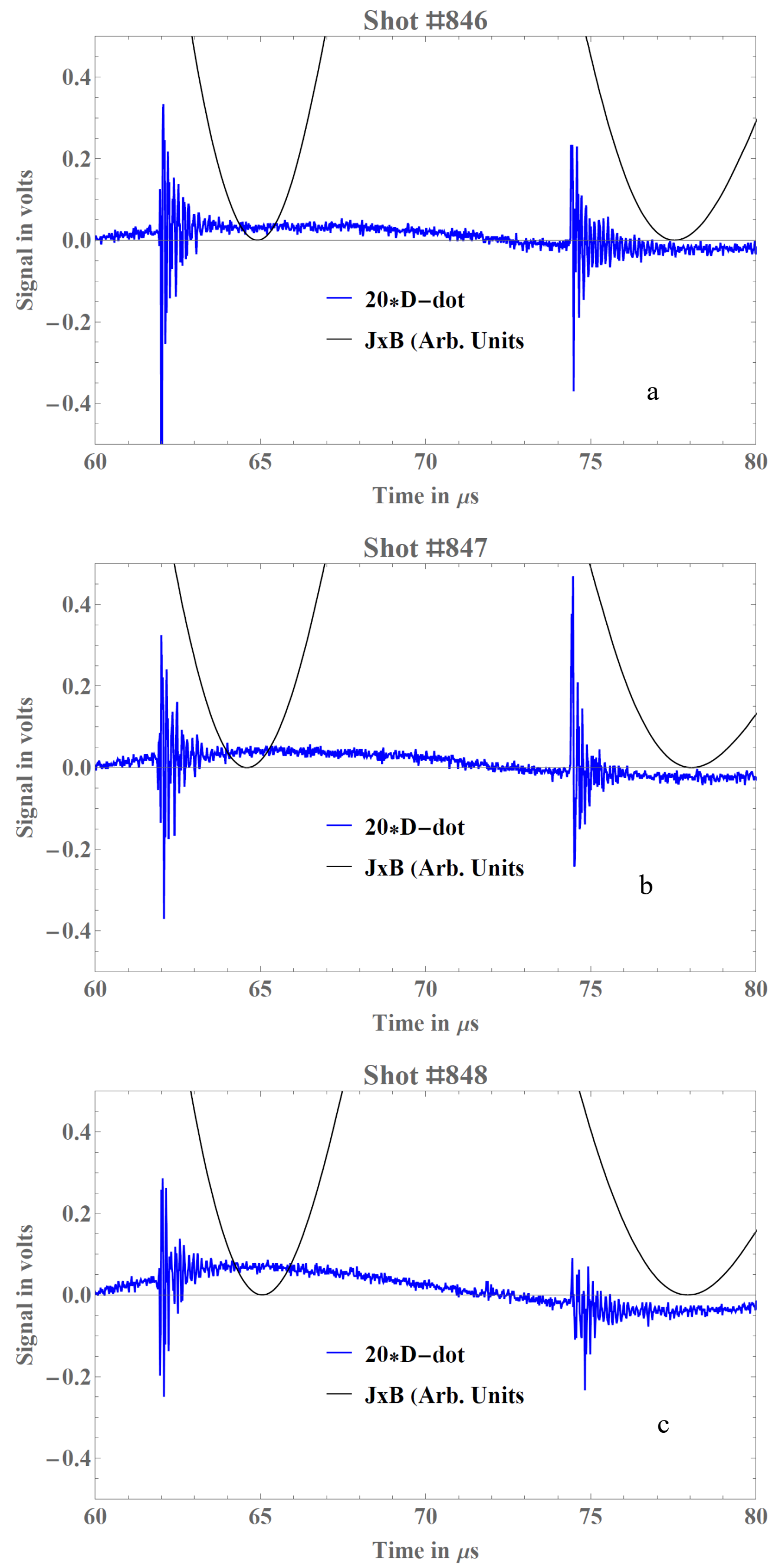

Fig-16: Relationship between the transients in the D-dot signal and the zeroes of the magnetic force density. The black curve is the square of the integration of the $B_\theta$-dot signal.

The transients are rapid damped oscillations. Their detailed structure differs considerably across the three shots. These transients are the principal evidence in support of the hypothesis described in the Introduction, as discussed later.

iii. Experiments near 100 torr of air, interelectrode gap ~ 3 mm, D-dot probe facing plasma

This series followed some changes in the probes. Both the magnetic and D-dot probes were moved closer to the device axis as shown in Fig 3. The D-dot probe was near the midplane of the interelectrode gap. Fig. 17 shows the results of 3 consecutive shots

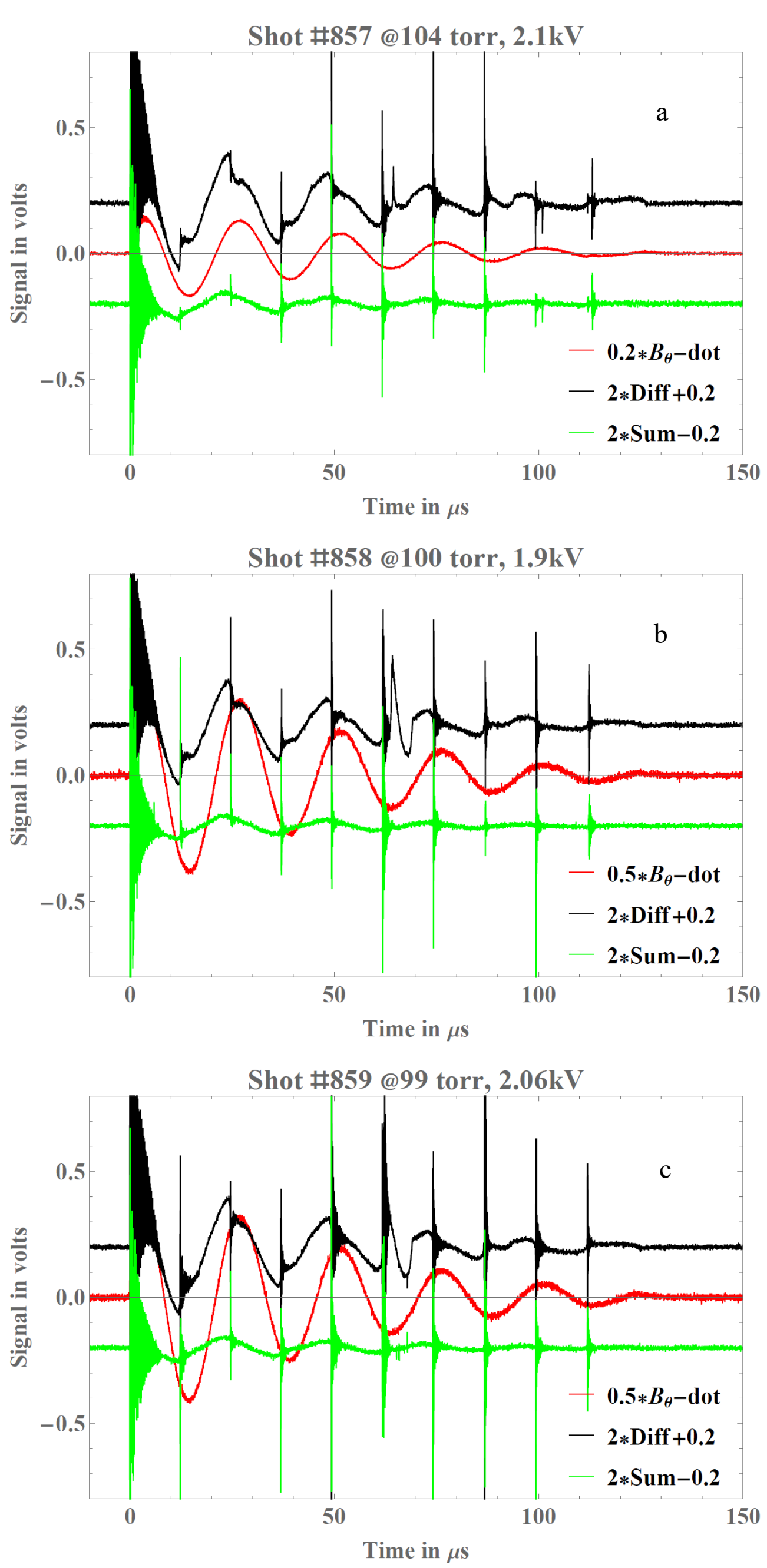

Fig-17. The Diff and Sum signals have significant transients just before the maxima of the current derivative. The current derivative signals have no corresponding transients.

Fig. 18 compares the Diff and D-dot signals with the integration of the $B_\theta$-dot probe signal.

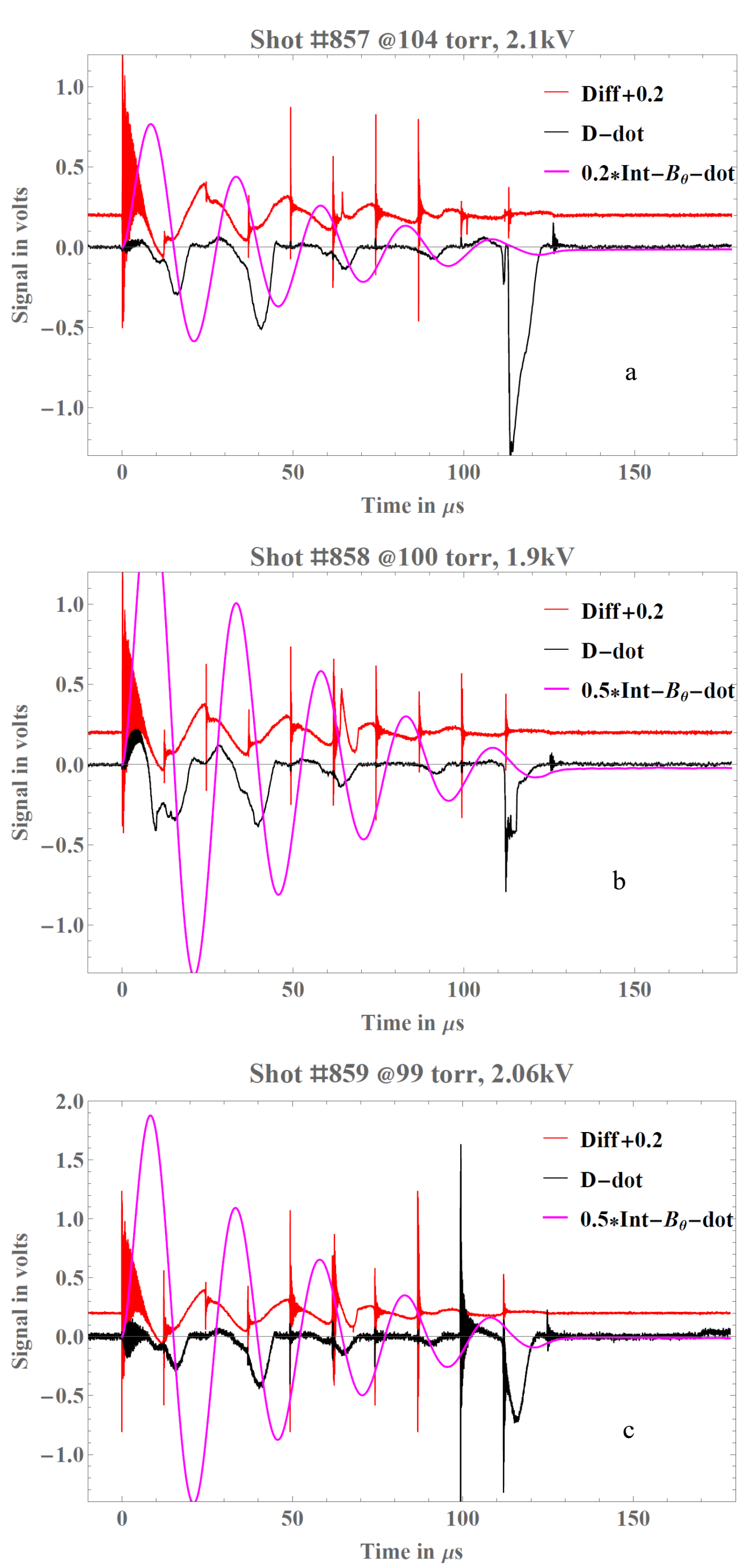

Fig-18: Comparison of Diff and D-dot signals with the integration of the $B_\theta$-dot signal. Note that the integration of the $B_\theta$-dot signal, proportional to the discharge current, recovers its baseline. Transients in the Diff and D-dot signals are seen to occur in the vicinity of the zeroes of the current.

Note that the broad negative excursions of the D-dot signals, possibly linked with radial expansion of the plasma near the current zeroes, apparently mask the sharp transients in shots #857 and #858 but they are clearly visible in shot #859.

iv. Experiments near and above 200 torr of air, interelectrode gap ~ 3 mm, D-dot probe facing plasma

This series was a continuation of the previous session, aimed at determining the effect of air pressure on the signals. The hardware and its deployment remained the same. The features described above are present in all the shots. Only one shot in this series is described to bring out some features not observed in the previous shots.

The first difference is seen in Fig 19, which shows that the magnetic probe signal has sharp transients and its integration does not recover its baseline as it does in Fig.-13. This suggests that some axial current continues to flow through the plasma after the capacitor has fully discharged, possibly indicating the presence of an autonomous dynamo. The early part of the integrated waveform is quite well fitted by the usual damped sinusoidal formula as shown in Fig 20

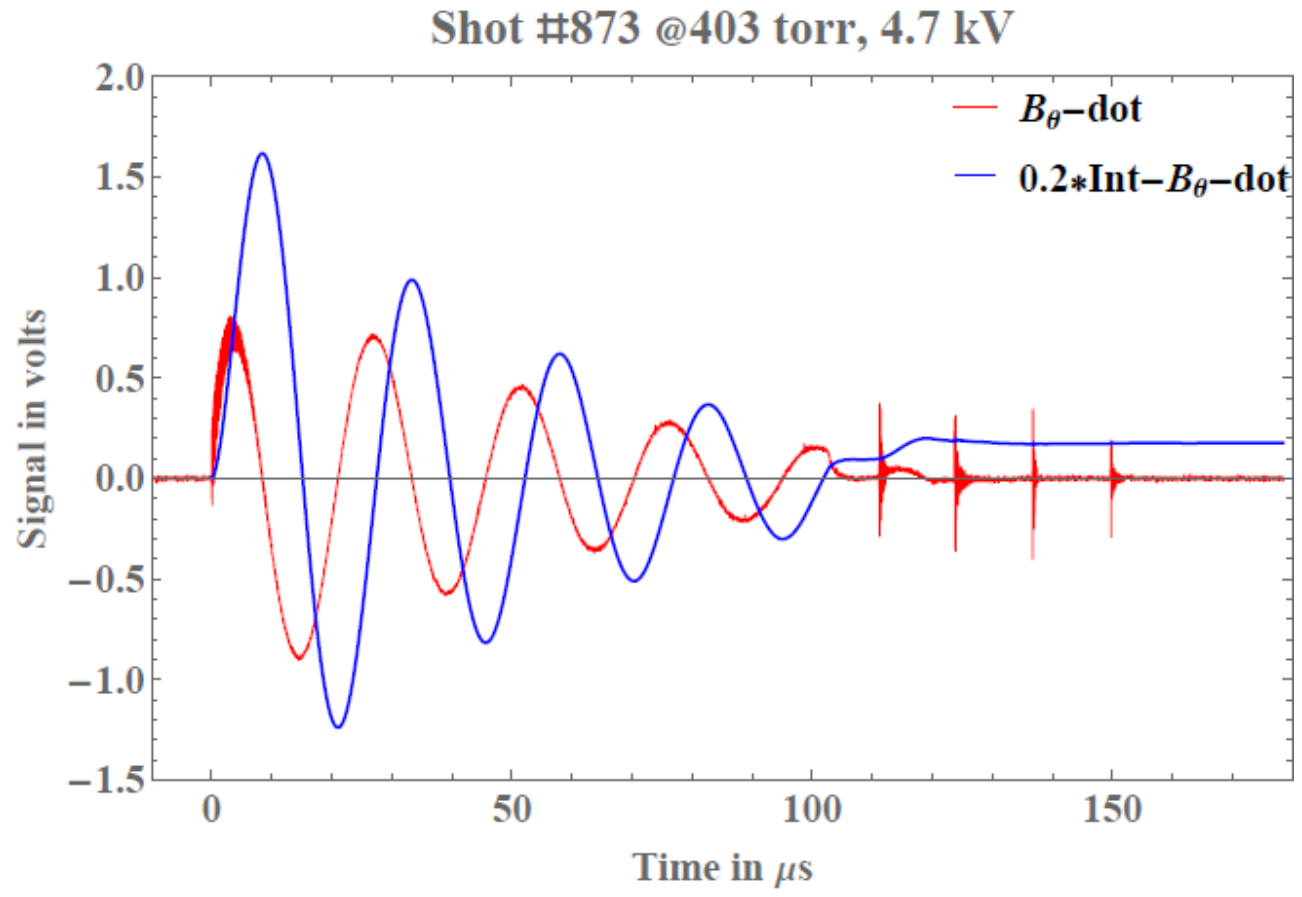

Fig.19: There are transients in the magnetic probe signal. As a result, its integration does not recover its baseline, suggesting the presence of axial plasma current flowing after the discharge of the capacitor.

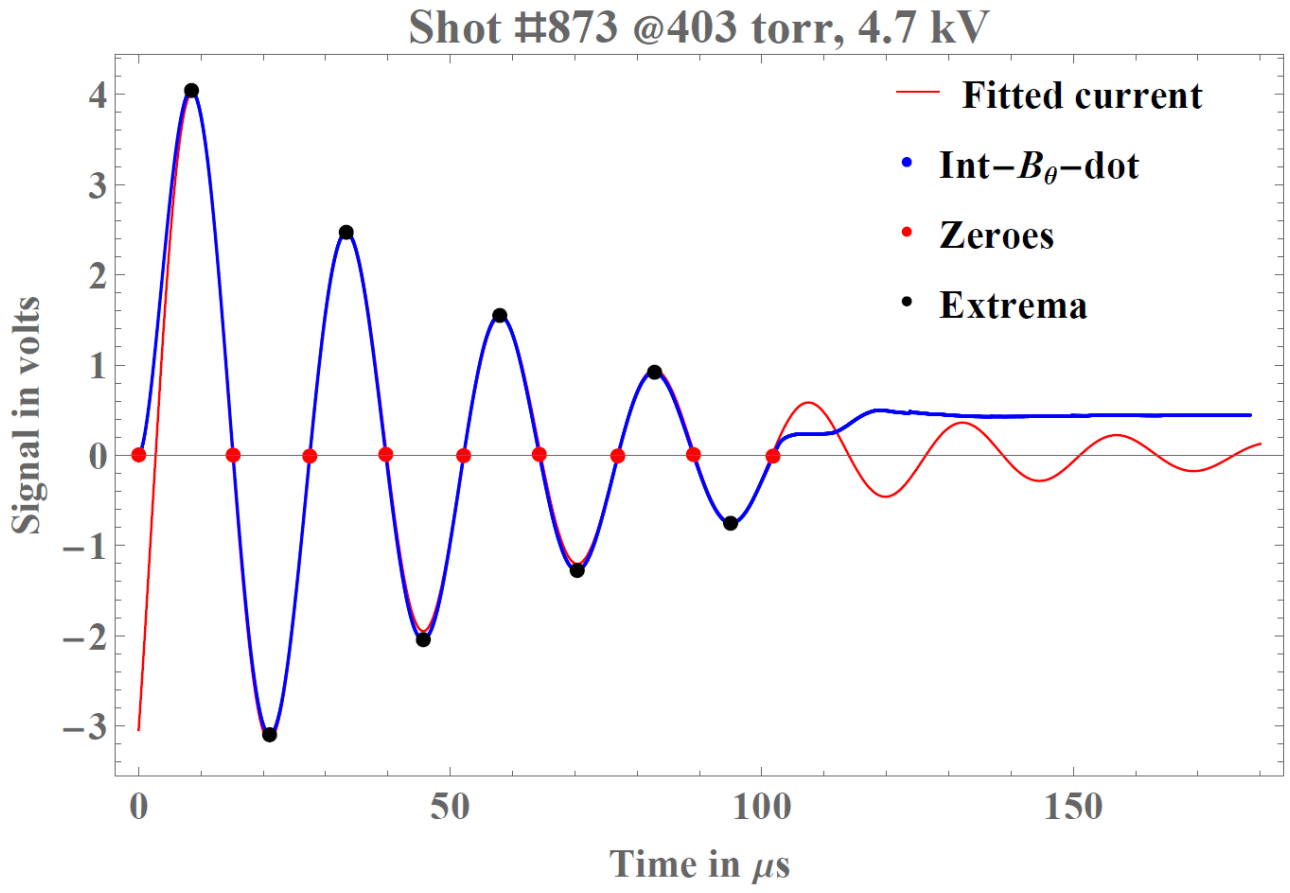

Fig-20: The initial portion of the integration of the magnetic probe signal is very well fitted by the usual damped sinusoidal expression.

The Diff and D-dot signals have prominent transients near the zeroes of the magnetic probe signal as seen in Fig.-21.

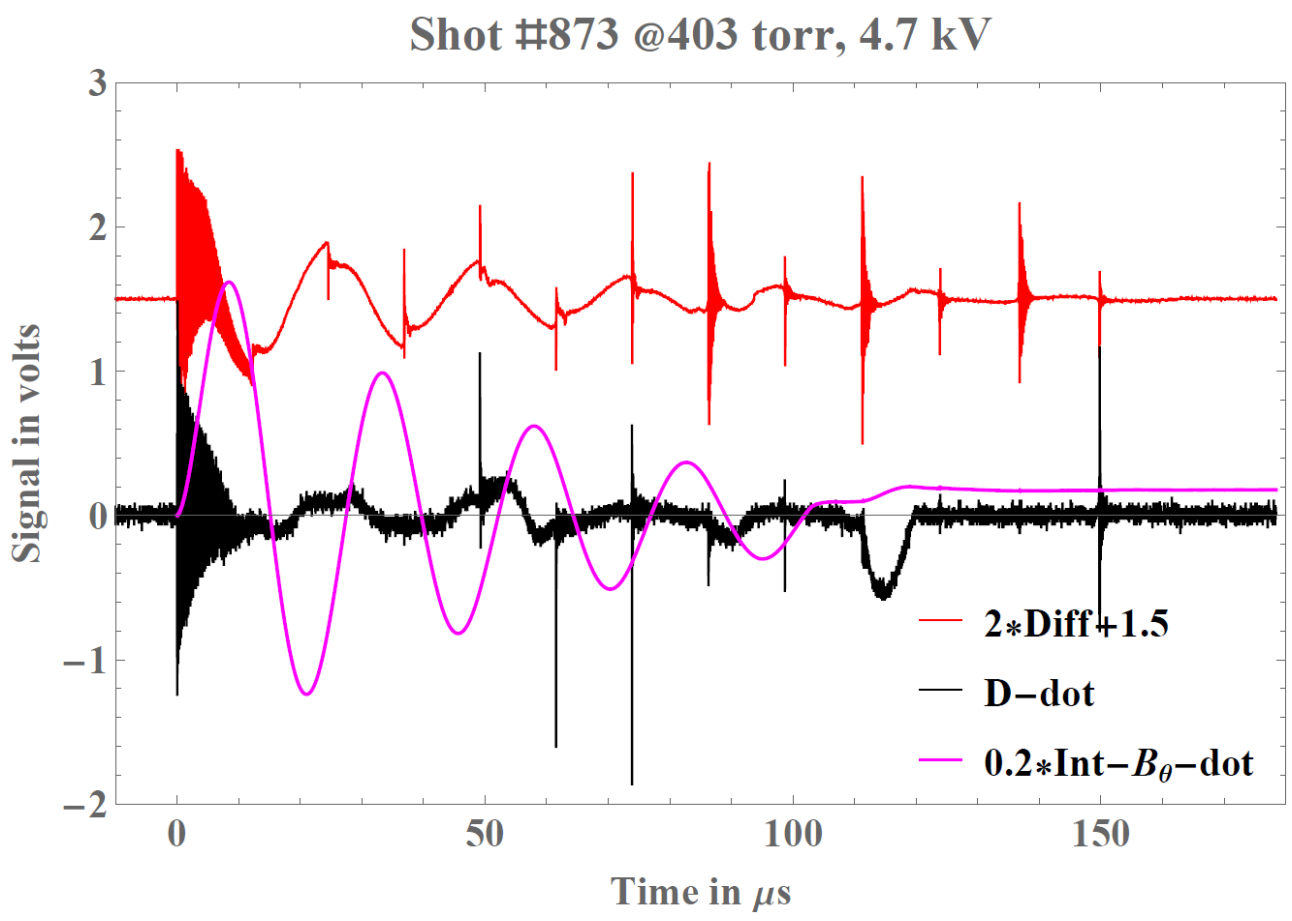

Fig-21. Diff and D-dot signals have prominent transients near the zeroes of the current

The integrations of Diff and Sum signals also persist after the discharge of the capacitor bank as seen in Fig 22.

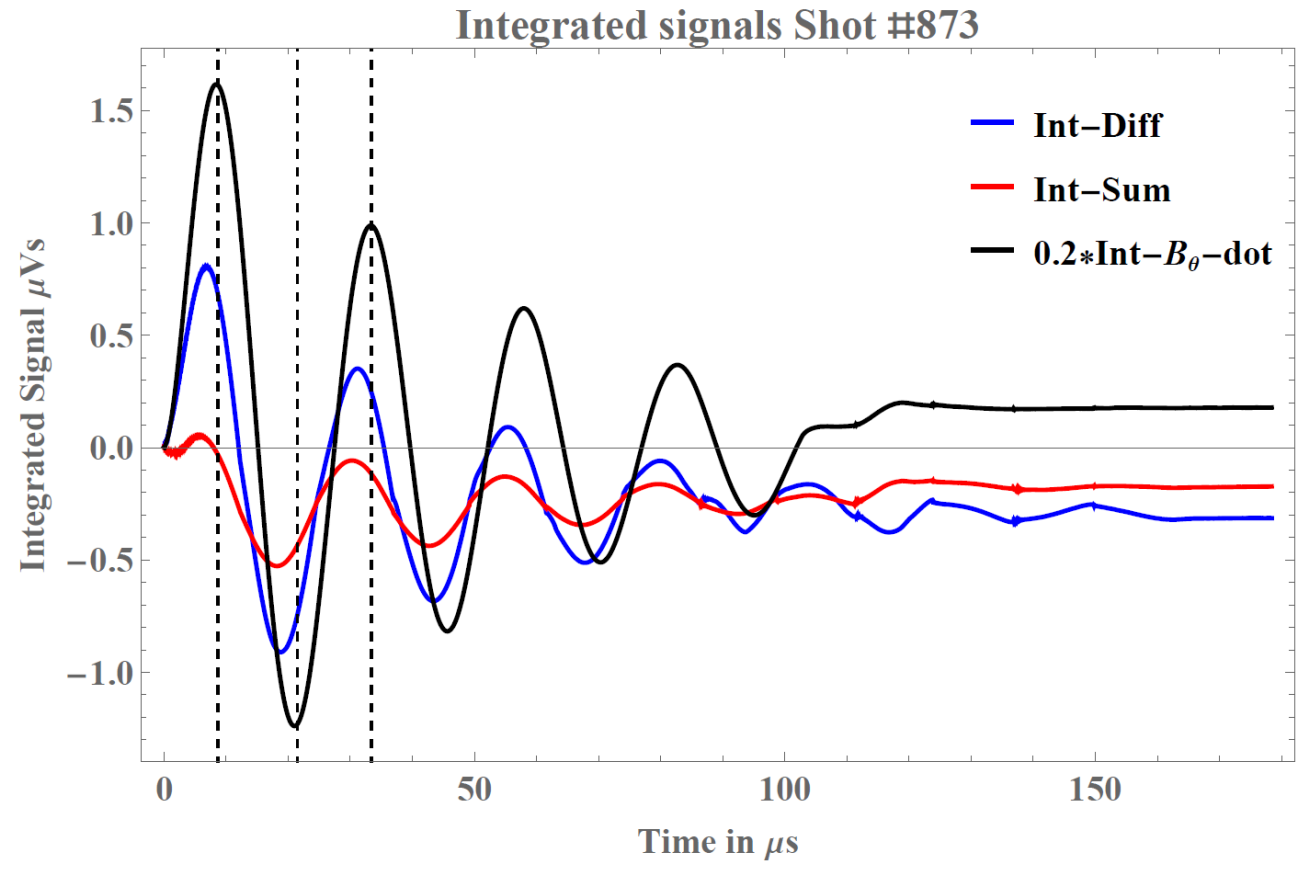

Fig-22: Integration of Diff and Sum signals persists beyond the cessation of the damped sinusoidal capacitor discharge waveform

v. Experiments near and above 200 torr of air, interelectrode gap ~ 3 mm, D-dot probe facing the solid electrode.

This series of experiments aimed to repeat the previous experiments with the D-dot probe facing the solid electrode so that its signal would be insensitive to the radial displacement of the plasma. Fig-23 shows that the D-dot signal becomes bidirectional as in Fig-11 where the D-dot probe faced a solid electrode and no longer has the broad negative peaks seen in Fig-8 and Fig-18 where it faced the plasma. This confirms our conjecture that the broad negative peaks in plasma-facing D-dot probes are because of plasma displacement while the damped oscillatory signal with overriding spikes is a feature of electrode-facing D-dot probes.

Fig-23 also shows that the magnetic probe signal has the typical damped oscillatory shape superimposed with transients, which are present near the maxima and more prominent near the end of the oscillatory waveform. The D-dot probe signal also a damped oscillatory shape with sharp spikes. The Diff and Sum signals have all the features already described above.

Fig. 24 shows the integrated signals. The integration of the magnetic probe signal is compared with a fitted damped sinusoidal waveform. The experimental curve stops oscillating and reaches an elevated baseline even as the fitted waveform continues to oscillate.

These results are discussed in Section IV.

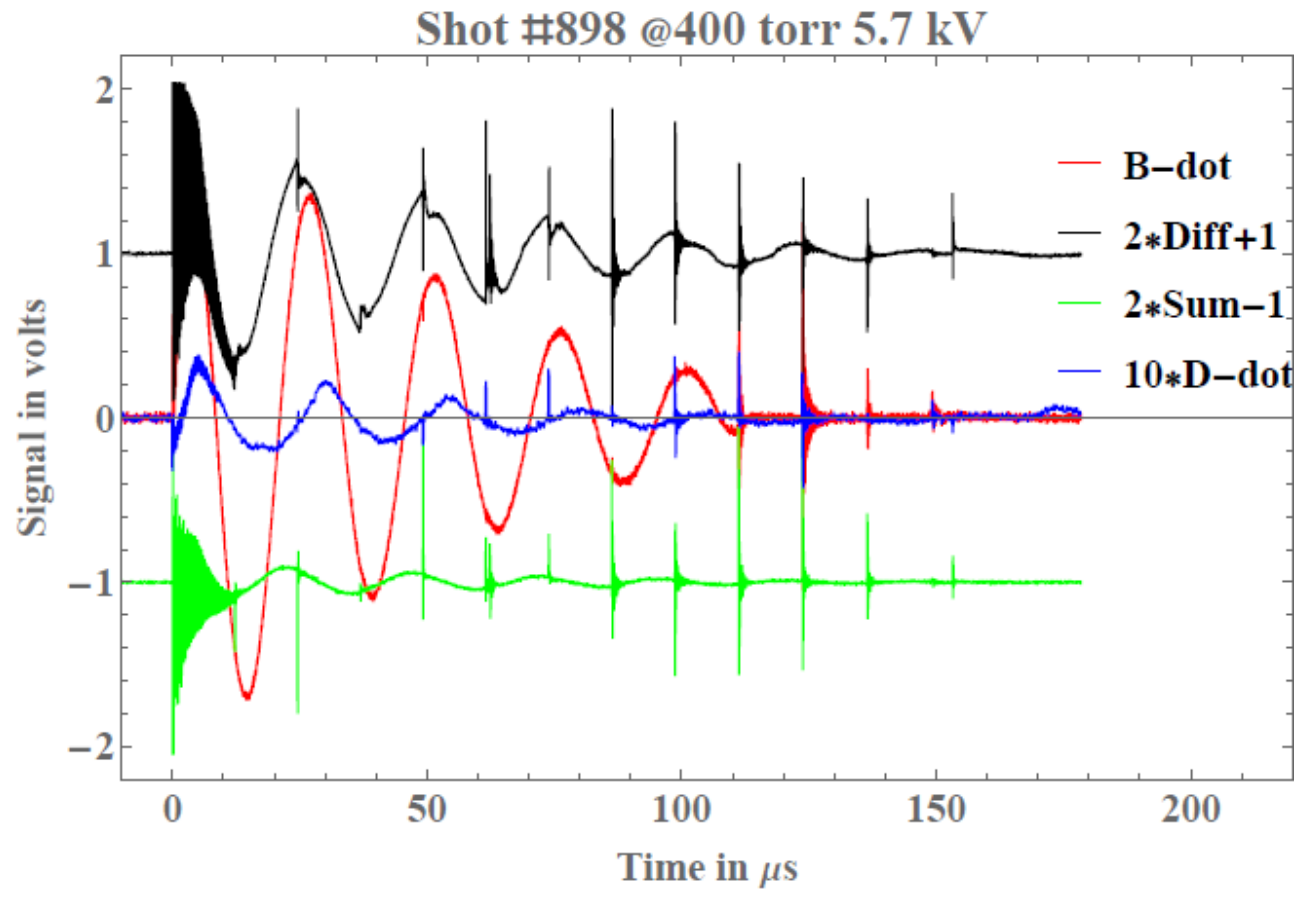

Fig-23: Data with electrode-facing D-dot probe.

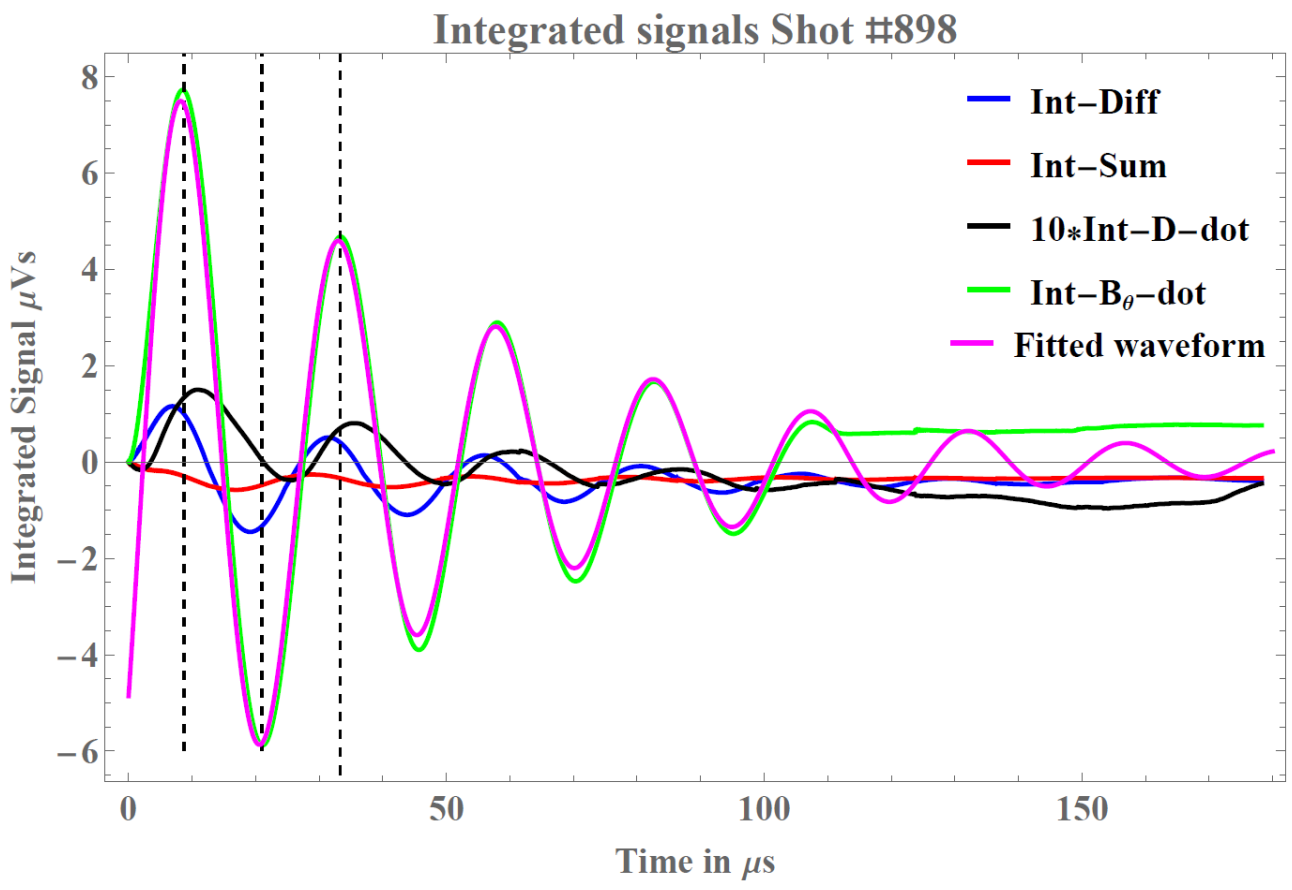

Fig-24: Integrated signals. A fitted damped sinusoidal waveform is displayed for comparison. Note that the fitted waveform continues to oscillate past 110 μs while the numerically integrated magnetic probe signal reaches a steady value much above the baseline. Similarly, integrations of the Diff, Sum and D-dot signals reach elevated baselines after the current stops oscillating.

## IV. Discussion:

The results presented above support the following inferences:

1) Emission of poloidal magnetic flux:

Signals derived from clockwise (CW) and counterclockwise (CCW) diamagnetic loops are almost, but not exact, mirror images of each other. These loops are very well centred with respect to the device axis and lie in planes perpendicular to it. They have identical dimensions to within manufacturing tolerances. The signals represent integration of the azimuthal component of the electric field (along the loop) with respect to the azimuthal angle. This signal can arise only if the loops enclose poloidal magnetic flux (PMF). Half the difference in the CW and CCW signals is numerically equal to the rate of change of the enclosed PMF. This conclusion directly follows from Faraday's Law and is beyond any doubt.

2) Emission of electric flux:

The fact that the CW and CCW signals are not exact mirror images of each other as expected from Faraday's Law is a reflection of the capacitive coupling between the power circuit driving the plasma and the measuring circuit connected to the oscilloscope. Electric flux emitted by the power circuit, which is proportional to the instantaneous electric charge on the high voltage electrode and the plasma, is incident on the CW and CCW diamagnetic loops, changing their potential with respect to the electrical ground of the oscilloscope. This effect is geometry dependent and is influenced by the dielectric environment such as the plastic hexagon providing mechanical support to the CW and CCW loops. Half the sum of the CW and CCW signals (referred to as the Sum signal) represents the effect of electric flux on the diamagnetic loops. The fact that the Sum signal resembles a damped oscillatory waveform corroborates this interpretation.

The D-dot probe provides an independent diagnostic related to the emission of electric flux. Its small active dimension enhances its sensitivity to localized changes in the geometry of the electric-charge-bearing environment, be it the high voltage electrode or the plasma. When the D-dot probe faces the solid electrode, this geometry is unchanging and hence the signal represents the dependence of the incident electric flux on circuit-determined changes in the instantaneous charge on the electrode, leading to resemblance with a damped oscillatory discharge. On the other hand, when it faces the plasma, it is additionally sensitive to its radial displacement. The broad negative peaks in the D-dot signals in Fig-7 and Fig-18 indicate that the plasma-facing probe is sensing radial displacement of the plasma occurring on a microsecond timescale. A negatively charged object approaching and receding from the probe would produce a similar effect.

3) Flux of the azimuthal magnetic field:

The magnetic probe shown in Fig-3 is sensitive to the flux of the azimuthal magnetic field passing through its turns, which is in turn proportional to the circuit-determined current flowing through the electrode and the plasma. The signal from the magnetic probe is proportional to the current derivative when the relative position between the magnetic probe and the current carrying element is constant.

For the case of 12 mm interelectrode gap, illustrated in Fig-4 to Fig-10, the magnetic probe signal has a strange but reproducible shape, not quite the damped oscillatory waveform expected from a capacitor discharge circuit. This probably reflects displacement of the plasma in a kink mode resulting in a time-varying inductance. Such displacement was considered an unwanted distraction and led to reduction of the interelectrode gap to 3 mm in subsequent experiments. The case of 12 mm interelectrode gap, reported above as an illustration of the rich phenomenology of a simple spark discharge, is omitted from further discussion because of this anomaly.

There is no capacitive coupling between this probe and the power circuit because the electric flux terminates on the insulated aluminium foil of its electrostatic shield and does not couple with the measurement circuit. This implies that observed features of the magnetic probe signal cannot be ascribed to the capacitive coupling. In contrast, the magnetic probe used in Ref [1] had no electrostatic shielding and it is quite likely that reported features of the observed signal [1] were a property of the capacitive coupling rather than the inductive coupling.

4) Damped oscillatory current flow:

The discharge of the capacitor through the air spark plasma is expected to create a damped sinusoidal current waveform, whose derivative is measured by the magnetic probe. This requires that the plasma does not introduce a significant time varying contribution to the inductance.

In many cases, this current waveform fits the theoretical expression quite well till the baseline is attained, as in Fig-13. Its oscillation period of 25 μs defines the reference timescale of this experiment. The oscillations continue up to 150 μs, giving 5 complete cycles. In a significant number of discharges, the magnetic probe signal, proportional to the current derivative, has multiple transients slightly before the zeroes of the current. As a result, the integration of the magnetic probe signal deviates from a damped sinusoidal waveform fitted to its early portions, attaining a non-oscillatory slowly varying shape. This is seen in Fig. 19, 20 and 24.

This behaviour is quite unexpected. Since the signal is shielded from the electric flux emitted by the power circuit, the observed transients must be produced by some plasma phenomenon that interferes with the axial transport of current by the electrons. In view of the fact that this is a collision-dominated partially ionized plasma and the electron current is resistively limited, this implies that it is *the axial electric field driving the electron drift that is exhibiting the transient behaviou*r. Implications of this inference are discussed later.

5) Transients in the Sum signal:

The Sum signal has a damped oscillatory waveform with sharp transients near the extrema. The Sum signals are quite reproducible as seen in Fig 8. The transients coincide with similar transients in the Diff signal and the D-dot signal.

6) Transients in the Diff signal:

The Diff signal, which is the rate of change of the poloidal magnetic flux passing through the diamagnetic loops, consistently exhibits pronounced transients at different pressures. Fig-14 and Fig-15 demonstrate that (1) the transients have a damped oscillatory waveform with a sub-microsecond period, much smaller than the 25-microsecond time period of the capacitor discharge; (2) a peak that can be nearly an order of magnitude higher than its baseline value just before the transient begins; (3) occurrence near the extrema of the magnetic probe signal, which are zeroes of the integrated magnetic probe signal proportional to the current.

7) Transients in the D-dot signal:

Transients in the D-dot signal are sometimes masked by the plasma displacement effect when the probe faces the plasma as shown in Fig 18 but are seen prominently near the end of the discharge. Fig-16 and Fig-23 illustrate that transients in the D-dot signal with probe facing the electrode have a damped oscillatory waveform on a sub-microsecond timescale and that they are coincident with transients in the Diff and Sum signals.

8) General nature of observed features:

The point about displaying data from three consecutive shots in four series is to bring out the fact that the features described above are largely reproducible, with perhaps some variation in details. They are thus not an artifact. Since the equipment is relatively simple, these experiments can be repeated in many laboratories to cross-check on our experiments. This repeatability and potential replicability is fundamental to the interpretation of these phenomena in terms of the hypothesis of constrained plasma dynamics discussed in Section V and its significance in plasma physics at large discussed in Section VI.

## V. Interpretation in terms of the hypothesis of constrained plasma dynamics:

A mathematical theory that can reproduce the observed features would be highly desirable. Development of such a theory would be much facilitated if a qualitative narrative consistent both with the observations and with fundamental physics could be constructed first. This section is an attempt in that direction. Such a narrative would be particularly important were these experiments were to be replicated with improved instrumentation in other laboratories.

The phenomenon begins with formation a spark [14] between the two metal electrodes across which a slowly increasing voltage is applied until breakdown. This establishes a conducting channel through which the capacitor can discharge and thereby drive a current. This conducting channel is a weakly ionized plasma. Electrons are its primary current carriers, whose mobility (and hence the plasma resistivity) is mainly determined by collisions with neutrals. Energy deposition within the plasma takes place when the electrons are accelerated by the electric field between collisions and lose their kinetic energy to the neutrals during collisions. The discharge current builds up to a value large enough for the magnetic force exerted on the electrons to pull them towards the axis. The electrons pull the ions along using the electrostatic force. The ions in turn drag the neutrals along by collisions. The net radial implosive motion of the plasma driven by the magnetic force density is thus impeded by the combined thermal force density of neutrals, ions and electrons.

This is a situation where charged particles are placed in a weak axial magnetic field (the geomagnetic field) and are acted upon by a centrally directed force. For electrons, this force is the resultant of the Lorentz force and collisional friction. For ions, it is the resultant of the electrostatic force (which makes them follow the electrons) and collisional friction. It can be shown [5] by quite a general argument that the charged particles must then move in orbits whose average azimuthal current is paramagnetic: it tends to reinforce the initial axial magnetic field. This azimuthal paramagnetic current exists only so long as the net forces acting on the charged particles are directed towards an axis and acts as a source of poloidal magnetic flux emission.

When the damped oscillatory capacitor discharge current is about to pass through a zero, the magnetic force density decreases and is no longer sufficient to overcome the force of collisional friction on an individual charged particle. The paramagnetic azimuthal current, which is the source of the poloidal magnetic flux, suddenly switches off. The PMF already created must then decay. But its decay is resisted by Lenz's Law which creates an azimuthal electric field that tries to drive a current that tends to maintain the flux. This azimuthal electric field accelerates electrons which creates an azimuthal electron current sufficient to maintain the PMF. But because of their inertia, the electrons continue to move further and try to overcompensate for the PMF. Lenz's Law then tries to brake the electrons, leading to oscillations of the azimuthal electric field which decay down by collisions. This explains the transients in the Diff signals occurring suddenly and repeatedly slightly before the zeroes of the current waveform.

The oscillating axial magnetic field and the azimuthal current density associated with the slower decay of the PMF create a radially directed oscillating Hall electric field. Its divergence represents an oscillating charge density which acts as the source of the electric flux emission detected by the Sum signal and the D-dot probe signal. This explains occurrence of the transients in these two signals in coincidence with the transients in the Diff signal.

Transients in the magnetic probe signal show that the axial transport of electrons is being hindered by some plasma process. From the above discussion, it is clear that the plasma channel is host to oscillations of charge density. This charge density modulates the axial electric field driving the axial transport of electrons. This explains occurrence of transients in the magnetic probe signal.

The occurrence of elevated baselines of the integrations of magnetic probe, D-dot and Diff signals towards the end of the discharge are the most important observations in our experiment. These are clearly not the artifacts of inadequate algorithms for baseline and time-zero corrections and numerical integrations since they do reproduce the theoretical damped sinusoidal waveform with zero baseline in some shots. They represent the creation of a long lasting non-oscillating axial plasma current, poloidal magnetic flux and electric flux. We speculate that these phenomena are signatures of an autonomous dynamo mechanism that stores electric energy, creates and sustains these fields. The observed transients, linked above to the electron inertia, clearly play a central role in this mechanism.

## VI. Implications:

The hypothesis of constrained plasma dynamics essentially highlights the role of normally neglected electron inertia effects on plasma dynamics. Its basic contention is that the truncated continuum models of plasma with ab initio neglect of electron inertia are of limited validity both mathematically and physically. This is because different terms of such truncated models represent diverse physical phenomena which can never truly ensure strict compliance with Newton's Second Law at all times under all circumstances. A concrete experimental counterexample has been described and interpreted invoking the role of electron inertia.

Electron inertia effects are manifested through nonlinear interaction between two kinds of electron modes: (1) electrostatic modes, where the restoring force involves irrotational (curl-free) electric field, whose divergence is a net charge density; (2) electromagnetic modes where the restoring force involves solenoidal (divergence-free) electric field, related to a current density. Both modes operate at the local plasma frequency. The electrostatic mode is excited by any change to the electron density or pressure and is normally present as a density fluctuation in the plasma.

The nonlinear interaction between these modes is of two kinds: (a) through the mechanism of the Hall effect, which is itself a manifestation of the Lorentz force on electrons; (b) through the mechanism of the electron momentum convection term $\left(\vec{v}_e \cdot \vec{\nabla}\right)\vec{v}_e$. Both essentially create fields at sum and difference frequencies of the coupled modes.

For example, an arbitrarily small but nonzero initial axial magnetic field at zero frequency couples with the current density of the radial electrostatic electron mode at the plasma frequency to create an azimuthal Hall electric field at the plasma frequency. This resonantly drives the azimuthal electromagnetic mode at the plasma frequency which creates a Hall electric field at zero frequency proportional to the initial axial magnetic field, leading to its growth at a timescale much larger than the plasma period.

The electrostatic mode at plasma frequency also creates a zero-frequency electrostatic field via the electron momentum convection term. This manifests itself as an elevated baseline on the integration of the D-dot signal when the probe is facing a solid electrode. A combination of zero-frequency electric and magnetic fields can drive steady $\vec{E} \times \vec{B}$ drift currents in the plasma, forming the basis of an axisymmetric autonomous dynamo mechanism.

These aspects are clearly seen in the reported experiment. But their significance goes much farther. *Every finite plasma that is created out of a no plasma state has such modes and there is no region of the universe that is completely devoid of some magnetic flux*. Both these assertions can be taken as axioms with universal validity until a counterexample is discovered. In *all* such cases, the electron modes must spontaneously generate and emit poloidal magnetic flux and electric flux along with associated sustained $\vec{E} \times \vec{B}$ drift currents. That would be a very general mechanism for the origin of the axial magnetic field in astrophysical jets [6,7].

This phenomenon involves interaction of processes occurring at widely different timescales, a situation where the current practice of summarily neglecting electron inertia terms in continuum models of the plasma needs to be replaced by the mathematical technique of multiple scale analysis [13] – a perturbation theory technique that would use a small parameter related to the electron mass. The recent observations [8] of amplification of a kilogauss level axial magnetic field of a neodymium magnet by 4 orders of magnitude in an experiment with femtosecond laser provides a strong motivation for such redevelopment of continuum plasma models.

Generation of toroidal mega-gauss magnetic fields by nonparallel gradients of electron density and temperature has been known for quite a long time. Add to that mega-gauss level poloidal magnetic fields created using the mechanism described above and one could generate a short-lived tokamak-like configuration at near solid density of fusion fuel. One could even conceptualise ignition of the proton-boron fuel in a near solid density tokamak as a merger of magnetic and inertial confinement schemes. As the preliminary experience [8,15] shows, bridging the gap between the electron plasma timescale and the hydrodynamic timescale is an extremely difficult computational task. Redevelopment of continuum models of plasma physics using the multiple scale analysis technique [13] would bring clarity to the study of such phenomena.

## VII. Summary and conclusions:

This paper deals with the traditional neglect of electron inertia in the widely used continuum models of plasma. The resulting truncated models have an inbuilt paradox where terms in the truncated equation of motion are seen to be governed by a diverse set of phenomena and are thus constrained to evolve at mismatched characteristic timescales putting strict universal compliance with Newton’s Second Law in jeopardy. This jeopardy is sought to be resolved by the hypothesis of constrained plasma dynamics, which stipulates that the usually neglected electron inertia terms step in and restore the deficit in momentum balance at a much finer timescale – that of the electron plasma period. This happens by the creation of a poloidal magnetic field, whose signatures are detectable outside the plasma. Microscopic mechanisms for this process already described earlier [3,4,5] are recapitulated.

An experiment is constructed and performed to deliberately create a situation where the momentum balance according to the truncated equation of motion must be violated. Diagnostic measurements include a pair of identical diamagnetic loops in clockwise and counterclockwise orientations, a D-dot probe and a magnetic probe. The results indicate that signatures expected according to the hypothesis of constrained dynamics are observed repeatedly.

This invites redevelopment of fundamental plasma physics without ab initio neglect of electron mass using multiple scale analysis. This exercise can unlock several unsolved problems and lay the foundations of novel approaches to controlled nuclear fusion.

Acknowledgements: This work was supported by the Faculty of Physics at the University of Sofia.